\documentclass[aps,jcp,preprint,superscriptaddress]{revtex4}

\usepackage[utf8]{inputenc}

\usepackage{amsmath,amssymb}
\usepackage{tabularx,booktabs,array,dcolumn}
\usepackage{setspace}
\usepackage{vmargin}
\usepackage{graphicx,psfrag,subfigure}
\usepackage{xcolor}
\usepackage{booktabs}
\usepackage{multirow}

\usepackage{marvosym}

\usepackage[all]{xy}

\usepackage{tikz}
\usetikzlibrary{calc}
\usetikzlibrary{patterns}

\usepackage{threeparttable}

\newcommand{\mx}[1]{\boldsymbol{#1}}
\newcommand{\bos}[1]{\boldsymbol{#1}}
\newcommand{\mr}[1]{\mathrm{#1}}

\newcommand{\cm}{cm$^{-1}$}

\def\som{Supplementary Information}

\def\Aonep{\mr{A}_1^+}

\def\Ep{\mr{E}^+}
\def\Fonep{\mr{F}_1^+}
\def\Ftwop{\mr{F}_2^+}
\def\Aonem{\mr{A}_1^-}
\def\Atwom{\mr{A}_2^-}
\def\Em{\mr{E}^-}
\def\Fonem{\mr{F}_1^-}
\def\Ftwom{\mr{F}_2^-}

\def\jmet{j^\text{M}}
\def\jwat{j^\text{W}}
\def\ka{k_\text{a}}
\def\kc{k_\text{c}}

\def\metwata{$\text{CH}_4\cdot\text{H}_2\text{O}$}
\def\metwatb{$\text{CH}_4\cdot\text{D}_2\text{O}$}
\def\metwatc{$\text{CD}_4\cdot\text{H}_2\text{O}$}
\def\metwatd{$\text{CD}_4\cdot\text{D}_2\text{O}$}

\def\RR{\text{RR}}
\def\CR{\text{CR}}

\def\Eint{E_\text{c}}

\def\jint{j}

\begin{document}

\title{%
Rovibrational quantum dynamical computations for deuterated isotopologues of the methane-water dimer
}

\author{J\'anos Sarka}
\affiliation{Institute of Chemistry, E\"otv\"os Lor\'and University, P\'azm\'any P\'eter s\'et\'any 1/A, Budapest, H-1117, Hungary}
\affiliation{MTA-ELTE Complex Chemical Systems Research Group, P.O. Box 32, H-1518 Budapest 112, Hungary}

\author{Attila G. Cs\'asz\'ar}
\affiliation{Institute of Chemistry, E\"otv\"os Lor\'and University, P\'azm\'any P\'eter s\'et\'any 1/A, Budapest, H-1117, Hungary}
\affiliation{MTA-ELTE Complex Chemical Systems Research Group, P.O. Box 32, H-1518 Budapest 112, Hungary}

\author{Edit M\'atyus}
\email{matyus@chem.elte.hu}
\affiliation{Institute of Chemistry, E\"otv\"os Lor\'and University, P\'azm\'any P\'eter s\'et\'any 1/A, Budapest, H-1117, Hungary}

\date{\today}

\begin{abstract}
\noindent
Rovibrational states of the four dimers formed by the light and the heavy isotopologues of the methane and the water molecules are computed using a potential energy surface taken from the literature. The general rovibrational energy-level pattern characteristic to all systems studied is analyzed employing two models of a dimer: the rigidly rotating complex and the coupled system of two rigidly rotating monomers. The rigid-rotor model highlights the presence of rovibrational sequences corresponding to formally negative rotational excitation energies, which is explained in terms of the coupled-rotors picture.
\end{abstract}

\maketitle

\clearpage
\section{Introduction}
\noindent
Weak molecular interactions are important in many areas of chemistry because they make it possible to break and form molecular assemblies including macromolecules like proteins and are involved in the self-organization of materials.
Weakly-bound systems exhibit several interesting and unusual spectroscopic features. High-resolution spectroscopic measurements, when compared with the results of quantum dynamical computations,  provide a stringent test of the quality of potentials modeling the interactions of the monomers in the complexes. 
For quantum dynamical computations weakly bound complexes have further interest because they are usually floppy, highly delocalized systems, which challenge our traditional tools and concepts of molecular structure and dynamics.

The methane-water dimer is a floppy, weakly bound complex. 
It was subjected to microwave \cite{SuFrLoKa94} and high-resolution far-infrared \cite{DoSa94} spectroscopic experiments some twenty years ago. 
The elucidation of the observed far-infrared transitions by means of 
quantum dynamical computations had to wait until last year \cite{SaCsAlWaMa16}.
The experimentally measured \cite{DoSa94} and the computed rovibrational transitions up to $J=2$ \cite{SaCsAlWaMa16}, using an accurate \emph{ab initio} intermolecular potential energy surface (PES) \cite{AOSz05}, are in excellent agreement, yielding deviations on the order of a few \cm.
The rovibrational computations reproduce fine details of and provide an assignment to many experimental features. Most interestingly, the experimentally observed reversed rovibrational sequences (whereby within the usual molecular picture one would assign a ``negative'' rotational energy to a vibrational state, a characteristic feature of certain floppy molecular systems \cite{FaSaCs14,SaCs16,WoMa15,ScJeSc15,WaCa16,ScJeSc16,ScJeSc17}) were also obtained in the computations.

Reference~\cite{SaCsAlWaMa16} left the explanation of these reversed sequences to further work suggesting that perhaps they could be better understood in terms of a coupled-rotor model rather than using the traditional picture of a rotating-vibrating molecule. The purpose of the present study is to work out this idea. 
In fact, we propose and describe a general algorithm to compute the similarity (overlap) of eigenstates of two coupled rotors with the full intermolecular wave function. This coupled-rotor decompositions (CRD) scheme is generally applicable to any system which can be partitioned into two subsystems. 
Through the application of the CRD scheme to 
all four possible combinations of the light and deuterated monomers, \metwata, \metwatb, \metwatc, and \metwatd, we gain a general overview about the properties of the methane-water dimer.

\section{Computational details \label{ch:comp}}
\noindent%
The quantum dynamical computations of the present study were carried out with the GENIUSH \cite{MaCzCs09,FaMaCs11} code, which was reviewed in Ref.~\cite{12CsFaSzMa} highlighting quantum chemical algorithms and computer codes of the fourth age of quantum chemistry. 
In GENIUSH, the matrix representation of the Hamiltonian is constructed in the discrete variable representation (DVR) \cite{00LiCa} and the kinetic energy operator is evaluated numerically during the course of the computation for the selected internal coordinates and body-fixed frame. An iterative Lanczos algorithm \cite{lanczos} is used to compute the required eigenvalues and eigenvectors. In the present work, our in-house Lanczos implementation was 
improved and optimized for the determination of a large number of eigenstates.
Over the years a number of tools have been developed supporting the analysis of the computed rovibrational wave functions \cite{NMD10,SaCs16}.

Computations with the GENIUSH code feature a series of floppy
molecular systems \cite{MaCzCs09,FaMaCs11,14FaMaCs,FaCsCz13,16SaLaBrCs,FaSaCs14,SaCs16,ArNOp}. For example, molecular systems with reversed rovibrational sequences (hence termed ``astructural molecules'') were studied in Refs.~\cite{FaSaCs14,SaCs16}. A molecular complex with almost fully decoupled degrees of freedom resulting in long-lived resonances was presented in Ref.~\cite{ArNOp}. Furthermore, the flexibility of the code makes it possible to model non-adiabatic effects in the rovibrational spectrum by using different rotational and vibrational masses \cite{MaSzCs14}.

Hereby we only shortly summarize the computational parameters and explain any difference compared to our earlier work \cite{SaCsAlWaMa16}. In the present study we computed the lowest-energy few dozens rovibrational eigenstates (150 states for $J=2$)
for the four isotopologues of the methane-water dimer with $J=0,1,$ and $2$ rotational angular momentum quantum numbers utilizing the QCHB15 PES \cite{QuCoHoBo15}
\footnote{%
The QCHB15 PES is based on 30467 ab initio interaction energies computed at CCSD(T)-F12b/haTZ (aug-cc-pVTZ for the carbon and oxygen atoms, cc-pVTZ for the hydrogen atoms) level of theory and is fitted with a root-mean-square error of 3.5~\cm.}.

In the quantum dynamical computations the methane and the water monomer structures are fixed and six active internal
coordinates describe the intermolecular dynamics. According to a recent study about the variation of the monomer structures due to non-covalent interactions \cite{RATuSc17},
this appears to be a fairly good approximation. The active internal coordinates, ($R,\cos\theta,\phi,\alpha,\cos\beta,\gamma$), and the embedding are identical to those of Ref.~\cite{SaCsAlWaMa16}, but in this work  we use the vibrationally averaged, $r_0$, structures of the monomers summarized in Table~\ref{tbl:struct}.
These values were made available for the H$_2$O and D$_2$O isotopologues in Ref.~\cite{09CzMaCs}, whereas for CH$_4$ and CD$_4$ we computed them as part of this study using the T8~force field \cite{ScPa01}.
The atomic masses used throughout this work are $m(\mr{H})=1.007\ 825$~u, $m(\mr{D})=2.014\ 102$~u, $m(\mr{C})=12$~u, and $m(\mr{O})=15.994\ 915$~u. 

The same DVRs were used for the active degrees of freedom as in Ref.~\cite{SaCsAlWaMa16}, which are listed in Table~\ref{tbl:comp_det} for the sake of completeness. The global minimum (GM) structure of the QCHB15 PES is also given in Table~\ref{tbl:comp_det}.
The corresponding $(A,B,C)/\text{cm}^{-1}$ rotational constants of the (near-symmetric top) dimers used for the wave function analysis are 
$(0.158421, 0.157281, 4.14248)$ for \metwata, 
$(0.149861, 0.147929, 3.56864)$ for \metwatb, 
$(0.138355, 0.137491, 2.28551)$ for \metwatc, and
$(0.130425, 0.128981, 2.09921)$ for \metwatd.

For the short and unambiguous referencing of the various isotopologues and their energy levels, we introduce the notation
\metwata\ (HH), \metwatb\ (HD), \metwatc\ (DH), and \metwatd\ (DD).
For the computed (ro)vibrational states of $XY$ ($X,Y=$~H, D) with $J=0,1$, and 2 we shall use the labels $XY$--J0.$n$, $XY$--J1.$n$, and $XY$--J2.$n$, respectively, where the integer $n$ enumerates all computed levels (including degeneracies) in an increasing energy order.

\begin{table}
  \caption{%
    Structural parameters and the corresponding rotational constants of the methane and the water molecules in their ground vibrational state 
    ($X=\text{O}$ or C and $Y=\text{H}$ or D).
    \label{tbl:struct}
  }
  \begin{tabular}{@{}c@{\ \ }r@{\ \ }r@{\ \ }r@{\ \ }r@{\ \ }r@{}}
    \hline\hline
    & 
    \multicolumn{1}{c}{CH$_4$} & 
    \multicolumn{1}{c}{CD$_4$} & & 
    \multicolumn{1}{c}{H$_2$O} & 
    \multicolumn{1}{c}{D$_2$O} \\ 
    \hline
    $\langle r_{XY}\rangle$ $[\text{\AA}]$ & 
      1.11002 & 1.10446 && 0.97565 & 0.97077 \\
    $\langle \alpha_{YXY}\rangle$ $[^\text{o}]$ & 
      109.471 & 109.471 && 104.430 & 104.408 \\\\
    $A$ $[\text{cm}^{-1}]$ & ~       & ~       && 9.17136 & 4.80367 \\    
    $B$ $[\text{cm}^{-1}]$ & 5.09073 & 2.57303 && 14.0667 & 7.11182 \\
    $C$ $[\text{cm}^{-1}]$ & ~       & ~       && 26.3538 & 14.8010 \\ \hline\hline
  \end{tabular}
\end{table}

\begin{table}
  \caption{%
    Internal coordinates (Coord.), discrete variable representations (DVR), number of grid points ($N$), and grid intervals used in GENIUSH to compute the rovibrational states of the methane-water dimers.\label{tbl:comp_det}}
  \begin{tabular}{@{}c@{\ \ \ }r@{\ \ \ } l@{\ \ \ }r@{\ \ \ }r@{}}
    \hline 
    \multicolumn{1}{c}{\raisebox{-0.30cm}{Coord.}} & 
    \multicolumn{1}{c}{\raisebox{-0.30cm}{Min.$^\text{a}$}} &
    \multicolumn{3}{c}{Nuclear motion computations} \\[-0.15cm] 
    \cline{3-5} \\[-0.25cm] 
    & &
    \multicolumn{1}{c}{DVR type$^\text{b}$} & 
    \multicolumn{1}{c}{$N$} & 
    \multicolumn{1}{c}{Grid interval} \\
    \hline
     $R$ [\AA]              &  3.447 & PO Laguerre & 15  & scaled to [2.5,6.0] \\
     $\theta$ [$^\text{o}$] & 117.72 & PO Legendre & 14  & scaled to [1,179] \\
     $\phi$   [$^\text{o}$] &  90.00 & Exponential & 15  & unscaled on [0,360) \\ 
     $\alpha$ [$^\text{o}$] & 297.65 & Exponential &  9  & unscaled on [0,360) \\ 
     $\beta$  [$^\text{o}$] & 112.80 & PO Legendre & 21  & scaled to [1,179] \\
     $\gamma$ [$^\text{o}$] & 293.52 & Exponential & 21  & unscaled on [0,360) \\ 
    \hline 
  \end{tabular}
\\[0.5cm]
\begin{flushleft}
     $^\text{a}$ %
     Equilibrium geometry in internal coordinates corresponding to the global minimum (GM) of the QCHB15 PES \cite{QuCoHoBo15}. \\
     $^\text{b}$ %
     PO: potential-optimized DVR (optimization of the DVR points for a 1-dimensional model) \cite{WeCa92,EcCl92}. \\
\end{flushleft}
\end{table}

\section{Assignment of the computed rovibrational states\label{ch:assign}}
\noindent%
The direct solution of the rovibrational Schr\"odinger equation provides numerically correct (``exact'') energy levels and wave functions, which carry detailed (``all'') information about the system. They can be used to make a direct comparison with experimentally observable transitions, to compute electric dipole transition intensities, etc. 
Although the wave functions carry ``all'' the information that can be learned about each state, it is usually overly complicated to ``read'' them and we may prefer to think about the states in terms of some well-established but simpler model. These simpler general descriptions include determination of the irreducible symmetry labels of the states based on the underlying molecular symmetry (MS) group, see, \emph{e.g.,} Refs.~\cite{BuJe98};
characterization of the nodal structure (``node counting''), see, \emph{e.g.,} Refs.~\cite{SaCs16,SaCsAlWaMa16,ArNOp};
computation of expectation values of structural parameters, see, \emph{e.g.,} Ref.~\cite{SaCsAlWaMa16};
or 
determination of the similarity (overlap) of the full wave function with wave functions of simple model systems as advocated, \emph{e.g.,} in Ref.~\cite{NMD10}.
For this purpose, the harmonic oscillator model of vibrational states and the rigid rotor (RR) model of rotational states are the working horses of molecular spectroscopy.

Another limiting model, relevant to weakly-bound dimers including the present one, is obtained by means of the coupling of rotating (rigid) subsystems. In what follows, we define 
an algorithm, which we call the coupled-rotor decomposition (CRD) scheme,
in order to compute the overlap of the full wave function (which might be available only in some grid representation, as it is the case in a GENIUSH computation) with the coupled-rotor functions.

\subsection{The coupled-rotor decomposition\label{ch:crd}}
\noindent%
The development and use of the CRD scheme has been motivated by the interesting energy-level structure of the methane-water dimer but it is useful for the understanding of the internal dynamics of other floppy systems which can be partitioned into two subsystems. 

In order to take a closer look at the model which we would like to use to characterize the full system, we reiterate that it consists of two rigid rotors fixed at a certain distance and the rotors are coupled to some total angular momentum state without being under the influence of any external potential. 
To assign the wave functions of the full system to states of the coupled, rotating monomers, we define a three-step numerical approach:
\begin{itemize}
 \item 
   Step~(1): Compute the rovibrational states 
of a 5-dimensional model using the same coordinate and grid representation as for the full model but with fixed intermonomer separation, $R$, and with $V=0$, \emph{i.e.,} without the influence of the PES. In this step the coupled-rotor (CR) eigenfunctions (and eigenvalues) are obtained in exactly the same grid representation for the five angles as in the full computation.
  \item
Step~(2): Calculate the eigenenergies of the 5D CR system with $J$ total angular momentum using the expressions derived for dimers in Ref.~\cite{BrAvSuTe83} corresponding to some monomer rotational states, $(j_A,j_B)$ and some coupling. The calculated CR energies are associated with angular momentum labels of the subsystems. 
  \item
Step~(3): 
By matching the energies of Steps~(1) and (2), the CR eigenfunctions in the desired representation computed in Step~(1) are associated with the monomer rotational angular momentum and coupling labels of Step~(2).
\end{itemize}

For Step~2, we reiterate the relevant expressions of Ref.~\cite{BrAvSuTe83}  to calculate the energy levels of an $A$--$B$ coupled-rotor system with $J$ total angular momentum quantum number: 
\begin{align}
  E^{(\CR)}(j_A,j_B,j,J)
  =
  E^{(\RR)}_{A}(j_A)
  +
  E^{(\RR)}_{B}(j_B)
  + 
  \Eint(j,J)
  \label{eq:ecpl}
\end{align}
where $j$ is the internal angular momentum quantum number, which results from the Clebsch--Gordan-type \cite{BuJe98} coupling of the angular momenta of $A$ and $B$, and it is thus restricted by the $|j_A-j_B| \leq j \leq |j_A+j_B|$ condition. 
$E^{(\RR)}_{A}(j_A)$ and 
$E^{(\RR)}_{B}(j_B)$ are the rigid-rotor energies of the monomers.  
$\Eint(j,J)$ can be obtained from diagonalizing the 
$\mx{K}\in\mathbb{R}^{\text{min}(j,J)\times\text{min}(j,J)}$ coupling matrix, 
the angular part of Eq.~(42) of Ref.~\cite{BrAvSuTe83}:
  \begin{align}
    K_{\Omega',\Omega}
    =
    \frac{\hbar^2}{2\mu R^2}
    &\left[%
      \delta_{\Omega',\Omega} (J(J+1)+j(j+1)-2\Omega^2)
    \right.
    \nonumber\\
    &+ 
    \left.
      \delta_{\Omega',\Omega+1} C_{j\Omega}^+ C_{J\Omega}^+ 
     +\delta_{\Omega',\Omega-1} C_{j\Omega}^- C_{J\Omega}^-
    \right] ,
    \label{eq:Kmx}
  \end{align} 
where $C_{jk}^{\pm}=\sqrt{j(j+1)-k(k\pm 1)}$,
$\Omega=-\text{min}(j,J),\ldots,\text{min}(j,J)$, and 
$\mu^{-1}=m_A^{-1}+m_B^{-1}$ is the reduced mass of the effective rotating diatom, which connects the centers of mass of the monomers.
$m_A$ and $m_B$ are the total mass of monomer $A$ and $B$, respectively, and
$\delta_{i,j}$ is the Kronecker delta function.
We mention that since the Hamiltonian of Ref.~\cite{BrAvSuTe83} does not contain any mixed radial-angular differential operators, the expressions for a fixed intermonomer distance, Eq.~(\ref{eq:Kmx}), can be obtained from  Eq.~(42) of Ref.~\cite{BrAvSuTe83} by simply retaining the angular part with some fixed $R$ value. 

The eigenvalues of the $\mx{K}$ coupling matrix, \emph{i.e.,} the coupling energies, disappear at infinite separation of $A$ and $B$. In this case the coupled-rotor energy, Eq.~(\ref{eq:ecpl}), is determined completely by the sum of the rigid-rotor energies of the monomers, which we call the free-rotor (FR) energy.
It is interesting to note that the eigenvalues of 
$\mx{K}$ 
can be obtained analytically. The analytic eigenvalues are provided for a few $J$ and $j$ values in the \som\ and the corresponding numerical value of the coupling energy is given for two intermonomer distances for all four studied isotopologues (Tables~S17--S19). 
The CR energy, $E^{(\CR)}(j_A,j_B,j,J)$ is
the sum of the FR energy (Tables~S13--S16) and the coupling energy (Tables~S17--S19).

As a result of Steps~1--3, the coupled-rotor functions, $\varphi^{(J)}(j_A,j_B,j)$ (with explicit $j_A, j_B, j$, and $J$ labels), are available in the same grid representation for the angular degrees of freedom as the full rovibrational wave function. In the next subsection we define the overlap of $\varphi^{(J)}(j_A,j_B,j)$ depending on five angles and the full wave function, $\Psi^{(J)}$, which depends on all six intermonomer coordinates.

\subsection{Numerical evaluation of the CRD overlaps\label{ch:crdnum}}
\noindent%
Since the 6-dimensional wave function is normalized to one, we can write
\begin{align}
 1
 &=
 \langle \Psi^{(J)}_{n} | \Psi^{(J)}_{n} \rangle_{R,\Omega} 
 \label{eq:overlap6D}\\
 &= 
 \sum_{k=-J}^J 
 \sum_{r=1}^{N_R} 
 \sum_{o=1}^{N_\Omega} %
  | {\tilde\Psi}_{n,k}^{(J)}(\rho_r,\bos{\omega}_o) |^2,   
 \label{eq:overlap6Db}
\end{align}
where
$\tilde\Psi$ refers 
to the DVR representation of the wave function with $\rho_r$ 
$(r=1,\ldots,N_R)$ grid points along the intermonomer distance, $R$,
and $\boldsymbol{\omega}_o$ $(o=1,\ldots,N_\Omega)$ grid points in the 5D angular subgrid, $\Omega=(\text{cos}\theta,\phi,\alpha,\text{cos}\beta,\gamma)$.
Next, introduce the truncated resolution of identity over some $M$ finite number of CR functions:
\begin{align}
  \mx{I}^{(J,\text{5D})}
  &\approx 
  \sum_{m=1}^M 
    | \varphi^{(J)}_{m} 
    \rangle_{\Omega}\ \cdot\ _{\Omega}\hspace{-0.04cm}\langle 
    \varphi^{(J)}_{m} |, 
  \label{eq:id5d}
\end{align}
which is inserted into Eqs.~(\ref{eq:overlap6D})--(\ref{eq:overlap6Db}):
\begin{align}
 1 &= \langle \Psi^{(J)}_{n} | \Psi^{(J)}_{n} \rangle \\
   &\approx
      \sum_{m=1}^M
        \Bigg\langle \Psi^{(J)}_n  | \varphi^{(J)}_m 
        \rangle_{\Omega}\ \cdot\ _{\Omega}\hspace{-0.04cm}\langle 
        \varphi^{(J)}_m | \Psi^{(J)}_n \Bigg\rangle_{R,\Omega} \\
   &=
   \sum_{m=1}^M
   \left(%
   \sum_{r=1}^{N_R}
     \left|
     \sum_{k=-J}^J 
     \sum_{o=1}^{N_\Omega}
       {\tilde\Psi}^{(J)}_{n,k}(\rho_r,\bos{\omega}_o) \cdot {\tilde\varphi}^{(J)}_{m,k}(\bos{\omega}_o)
     \right|^2
   \right).
   \label{eq:crd0} 
\end{align} 
Then, we define the $(n,m)$th element of what we call the CRD matrix 
for the $J$ total rotational quantum number as
\begin{align}
  \text{CRD}^{(J)}_{nm} 
  =
   \sum_{r=1}^{N_R}
   \left|%
   \sum_{k=-J}^J 
   \sum_{o=1}^{N_\Omega}
     {\tilde\Psi}^{(J)}_{n,k}(\rho_r,\bos{\omega}_o) \cdot {\tilde\varphi}^{(J)}_{m,k}(\bos{\omega}_o)
   \right|^2,
  \label{eq:defcrd}
\end{align}
where
$\Psi^{(J)}_n$ is the $n$th state and 
$\varphi^{(J)}_m$ is the $m$th CR function.

There are a few important properties of the CRD matrix which immediately follow from Eq.~(\ref{eq:defcrd}):
\begin{itemize}
\item[(a)] 
   The sum of the elements in each row, \emph{i.e.,} the CRD coefficients corresponding to some $\Psi^{(J)}_n$ state,
   is 1 for infinitely many CR functions (as the truncated resolution 
   of identity in Eq.~(\ref{eq:id5d}) becomes exact).
\item[(b)]  
  The sum of the elements in each column can be larger than $1$, because the same CR state can have a dominant contribution 
  to several $\Psi^{(J)}_n$ states. For the methane-water dimer, 
  the stretching excitations along the intermonomer separation, enumerated with $v=0,1,2,\ldots$, can have similar angular parts.
\item[(c)]
   The angular momentum coupling rules often allow, depending on the angular momenta of the full and the subsystems, a large number of CR functions with the same $(j_A,j_B)$ monomer rotational excitations. The individual contributions of the CR functions might be modest but the full $(j_A,j_B)$ block can have a substantial contribution to the $\Psi^{(J)}$ state. Therefore, the final assessment of the contribution (or dominance of some monomer excitation) $(j_A,j_B)$ must be based on the sum of the CRD coefficients corresponding to all $(j_A,j_B)$ monomer states.
\item[(d)]
   If the CR functions are determined with monomers infinitely far in space (in practice fixed at some large $R$ value), then the CR energy is equal to the FR energy, \emph{i.e.,} the simple sum of the RR energies (Tables~S13--S16). This limit facilitates the assignment of the monomer rotational states, $(j_A,j_B)$, but in this case the coupling and the quantum number corresponding to the rotation of an effective diatom connecting the centers of mass of the monomers cannot be determined.
\end{itemize}

\section{Symmetry of the methane-water coupled-rotor functions \label{ch:crsymm}}
\noindent%
While Section~\ref{ch:crd} applies to any system which can be partitioned into two subsystems, the symmetry analysis of this section applies only to the methane-water dimer. Nevertheless, a similar symmetry analysis can be carried out for other dimers as well.

In this section the characters of the methane-water CR representation are calculated and the representation is decomposed into a direct sum of irreducible representations (irrep decomposition) in the $G_{48}$ group \cite{DoSa94,SaCsAlWaMa16}. For convenience, the symmetry properties of the methane and the water monomer rotor functions are presented first within the $T_\text{d}$(M) and the $C_{\text{2v}}$(M) groups, respectively, which is followed by the symmetry analysis of the coupled-rotor functions of the dimer within the $G_{48}$ group. (For completeness, the character tables are reproduced in the \som.)

\paragraph{Methane monomer} 
Each element of a class has the same character, so we select one operator from each class and determine its character. The spatial rotation equivalent to the permutation-inversion operator is first written in the angle-axis parameterization (Figure~S1 of Ref.~\cite{SaCsAlWaMa16} visualizes the effect of the symmetry operations). Then, the Euler angles, $a,b,$ and $c$ (using the $z-x-z$ and active convention) are determined from the rotation matrix evaluated with Rodriguez' rotation formula in the angle-axis parameterization. 
For the symmetry operations (one from each class of $T_\text{d}$(M)) the $(a,b,c)$ Euler angles are as follows: 
    $(123)$: $(0,0,2\pi/3)$; 
    $(12)(34)$: $(\pi/3,\pi,0)$; 
    $[(1234)]^\ast$: $(\pi/6,\pi/2,-\pi/6)$; and 
    $[(34)]^\ast$: $(0,\pi,0)$.
Symmetric (as well as spherical) top rotational functions transform under the action of a rotation operator as \cite{Ho62,Ho63,AlLeCaPe11}:
\begin{align}
  \hat{R}(a,b,c)
  \varphi_{km}^{\jmet}
  =
  \sum_{k'=-\jmet}^{\jmet}
    D_{kk'}^{\jmet}(a,b,c) 
    \varphi_{km}^{\jmet} ,
\end{align}
where $D_{km}^{j}$ is the $(k,m)$th element of Wigner's D matrix corresponding to the $j$th irrep of the rotation group, and M denotes the methane monomer of the dimer.
Then, an operator $\hat{o}\in T_{\text{d}}(\text{M})$, for which $\hat{R}(a_o,b_o,c_o)$ is the equivalent rotation,
has the following character:
\begin{align}
  \chi^{\jmet}
  &=
  \sum_{k=-\jmet}^{\jmet}
    \langle %
      \varphi_{km}^{\jmet}
      |\hat{o}|
      \varphi_{km}^{\jmet}
    \rangle
  \nonumber\\
  &=
  \sum_{k=-\jmet}^{\jmet}
    D_{kk}^{\jmet}(a_o,b_o,c_o) 
\end{align}
where we also took advantage of the fact that the rotational functions are orthonormalized.

Using these properties the characters of a representation spanned by 
the $\lbrace \varphi_{km}^{\jmet}, (k=-\jmet,\ldots,\jmet) \rbrace$ 
functions are easily determined. 
The general expressions and the numerical values for $\jmet=0,1,\ldots,5$ are collected in Table~\ref{tab:charmet} (and the table can be extended straightforwardly to higher $\jmet$ values).
The last column of Table~\ref{tab:charmet} gives the irrep decomposition for the numerical examples, which is in agreement with the irrep decomposition of the methane rotational states in the $T(\text{M})$ group \cite{OhEn90,Hu90} and reproduces the results of Ref.~\cite{RaIbHo94} for the $T_\text{d}(\text{M})$ group. 

\paragraph{Water monomer} Table~\ref{tab:charwat} lists the general expressions \cite{BuJe98} for the characters of the $\jwat_{\ka\kc}$ rotational states of water (W) in the $C_{2v}$(M) group.

\paragraph{Coupled-rotor ansatz} 
The coupled-rotor ansatz can be understood as coupling 
(a) the symmetric top eigenfunctions of methane, $\varphi_{km}^{\jmet}$; 
(b) the asymmetric top eigenfunctions of water, 
$\phi_{\ka\kc,m}^{\jwat}$, 
and 
(c) the eigenfunctions of an effective diatomic rotor, $Y_{\Lambda}^m$, which corresponds to the rotation of the displacement vector connecting the centers of mass of the monomers. 
We denote this coupling scheme by
$[[\jmet,\jwat_{\ka\kc}]_j,\Lambda]_{JM}$, which corresponds to the function:
\begin{align}
  \Psi_{J}^M(\jmet_k,\jwat_{\ka,\kc},j,\Lambda)
  &=
  \sum_{\mu\in S_1}
    \langle
      \jint \mu, \Lambda(M-\mu) | JM
    \rangle\ 
    \psi_{\jint}^{\mu}     
    Y_\Lambda^{M-\mu}
 \nonumber \\
 &=
 \sum_{\mu\in S_1}
   \langle
     \jint \mu, \Lambda (M-\mu) | JM
   \rangle\ 
   Y_\Lambda^{M-\mu} \times
   \nonumber \\
   &\quad\quad\sum_{m\in S_2}
     \langle
       \jmet m ,\jwat (\mu-m) | \jint\mu
     \rangle\  
     \varphi_{km}^{\jmet}
     \phi_{\ka\kc,\mu-m}^{\jwat}
     \nonumber \\
  &\quad\quad\quad\text{with}\quad
  S_1=
   \lbrace%
    \mu\in\mathbb{N}_0: |\mu|\leq\jint\ \text{and}\ |M-\mu|\leq\  \Lambda %
  \rbrace, 
  \nonumber\\
  &\quad\quad\quad\quad\quad\quad
  S_2=
  \lbrace%
    m\in\mathbb{N}_0: |m|\leq \jmet\ \text{and}\ |\mu-m|\leq\jwat
  \rbrace ,
  \nonumber\\
  &\quad\quad\quad\quad\quad\quad\quad
  k=-\jmet,\ldots,\jmet .
  \label{eq:cransatz}
\end{align}
The label
$j$ is for the internal angular momentum quantum number
resulting from the coupling of the methane and the water rotors. 

An operator $\hat{O}\in G_{48}$ can be written as the product of $\hat{o}\in T_{\text{d}}(\text{M})$, methane permutation-inversion, 
$\hat{p}\in C_{\text{i}}=\lbrace E,E^\ast \rbrace$ acting on the effective diatom, and $\hat{q}\in C_{\text{2v}}(\text{M})$ water permutation-inversion operators. The character of $\hat{O}=\hat{o}\hat{p}\hat{q}$ is obtained as the trace of the matrix representation of $\hat{O}$ over the ansatz defined in Eq.~(\ref{eq:cransatz}):
\begin{align}
  \chi(\hat{O})
  &=
  \sum_{k=-\jmet}^{\jmet}
    \langle 
      \Psi_{J}^M(\jmet_k,\jwat_{\ka\kc},j,\Lambda)
      | \hat{O} 
      \Psi_{J}^M(\jmet_k,\jwat_{\ka\kc},j,\Lambda)
    \rangle
    \nonumber \\
  &=
  \sum_{k=-\jmet}^{\jmet}
     \sum_{\mu'\in S'_1}
     \sum_{m'\in S'_2}
     \sum_{\mu\in S_1}
     \sum_{m\in S_2}
  \nonumber \\
  &\quad\quad\quad\quad\quad\quad
      \langle
        j\mu',\Lambda(M-\mu') | JM
      \rangle
      \langle
        \jmet m', \jwat (\mu'-m')) | j\mu'
      \rangle 
  \nonumber \\
  &\quad\quad\quad\quad\quad\quad
      \langle
        j\mu,\Lambda(M-\mu) | JM
      \rangle
      \langle
        \jmet m, \jwat (\mu-m)) | j\mu
      \rangle 
    \nonumber\\
  &\quad\quad\quad\quad\quad\quad
    \langle %
      Y_\Lambda^{M-\mu'}|\hat{p}|Y_\Lambda^{M-\mu}
    \rangle
    \langle %
      \varphi_{km'}^{\jmet}
      |\hat{o}|
      \varphi_{km}^{\jmet}
    \rangle
    \langle %
      \phi^{\jwat}_{\ka\kc,\mu'-m'}
      |\hat{q}|
      \phi^{\jwat}_{\ka\kc,\mu-m}
    \rangle    
  \nonumber \\
  &=
  \sum_{k=-\jmet}^{\jmet}
     \sum_{\mu'\in S'_1}
     \sum_{m'\in S'_2}
     \sum_{\mu\in S_1}
     \sum_{m\in S_2}
  \nonumber \\
  &\quad\quad\quad\quad\quad\quad
      \langle
        j\mu',\Lambda(M-\mu') | JM
      \rangle
      \langle
        \jmet m', \jwat (\mu'-m')) | j\mu'
      \rangle 
  \nonumber \\
  &\quad\quad\quad\quad\quad\quad
      \langle
        j\mu,\Lambda(M-\mu) | JM
      \rangle
      \langle
        \jmet m, \jwat (\mu-m)) | j\mu
      \rangle 
    \nonumber\\
  &\quad\quad\quad\quad\quad\quad
    (-1)^{p\Lambda} \delta_{\mu'\mu}
    D^{\jmet}_{kk}(a_o,b_o,c_o) \delta_{m'm}
    \Gamma^{\jwat}_{\ka\kc}(\hat{q}) \delta_{\mu'-m',\mu-m}
  \nonumber \\
  &=
  \sum_{\mu\in S_1}
  \sum_{m\in S_2}
      \langle
        j\mu,\Lambda(M-\mu) | JM
      \rangle^2
      \langle
        \jmet m, \jwat (\mu-m)) | j\mu
      \rangle^2 
    \nonumber\\
  &\quad\quad\quad\quad\quad\quad
    (-1)^{p\Lambda} 
    \left[%
      \sum_{k=-\jmet}^{\jmet}
        D^{\jmet}_{kk}(a_o,b_o,c_o)
    \right]
    \chi^{\jwat}_{\ka\kc}(\hat{q})
  \nonumber \\
  &=
    \chi^\Lambda(\hat{p})\ 
    \chi^{\jmet}(\hat{o})\ 
    \chi^{\jwat}_{\ka\kc}(\hat{q})
  \label{eq:char}
\end{align}
where $\chi^\Lambda(\hat{p})=(-1)^{p\Lambda}$ is the character of the effective diatomic rotation, with $p=1$ for inversion and $p=0$ otherwise. 
$\chi^{\jmet}(\hat{o})$ and $\chi^{\jwat}_{\ka\kc}(\hat{q})$
are the characters of the permutation-inversion operations of the methane and water molecules 
listed in Tables~\ref{tab:charmet} and \ref{tab:charwat}, respectively.

The symmetry rules (Table~\ref{tab:crsymmj0} and Tables~S23--S26) derived in this section together with the numerically computed CRD tables make it possible to carry out  the symmetry assignment of the variationally computed rovibrational states in an automated fashion. 

\begin{table}
  \caption{%
    Characters and irrep decomposition 
    of the rotational functions of methane
    in the $T_\text{d}$(M) molecular symmetry group.
    \label{tab:charmet}
  }
\scalebox{0.9}{%
  \begin{tabular}{@{}l r@{}l r@{}l r@{}l r@{}l r@{}l c@{}}
  \cline{1-12} \\[-0.4cm]
  \cline{1-12} \\[-0.4cm]
    & 
    \multicolumn{2}{c}{$E$} & 
    \multicolumn{2}{c}{(123)} & 
    \multicolumn{2}{c}{(14)(23)} & 
    \multicolumn{2}{c}{[(1423)]$^\ast$} & 
    \multicolumn{2}{c}{[(23)]$^\ast$} & Irreps \\
     & 
    \multicolumn{2}{c}{1} & 
    \multicolumn{2}{c}{8} & 
    \multicolumn{2}{c}{3} & 
    \multicolumn{2}{c}{6} & 
    \multicolumn{2}{c}{6} & \\        
  \cline{1-12} \\[-0.4cm]
    $\jmet_k$
    & 
    \multicolumn{2}{c}{$2\jmet+1$} & 
    \multicolumn{2}{c}{%
\tiny
$\sum\limits_{m=-\jmet}^{\jmet} D^{\jmet\ast}_{m,m}(0,0,\frac{2\pi}{3})$
} & 
    \multicolumn{2}{c}{%
\tiny
$\sum\limits_{m=-\jmet}^{\jmet} D^{\jmet\ast}_{m,m}(\frac{\pi}{3},\pi,0)$
} & 
    \multicolumn{2}{c}{%
\tiny
      $\sum\limits_{m=-\jmet}^{\jmet} D^{\jmet\ast}_{m,m}(\frac{\pi}{6},\frac{\pi}{2},-\frac{\pi}{6})$
} & 
    \multicolumn{2}{c}{%
\tiny
$\sum\limits_{m=-\jmet}^{\jmet} D^{\jmet\ast}_{m,m}(0,\pi,0)$
} & \\
\cline{1-12} \\[-0.4cm]
    0 & 
   &1 &   
   &1 &
   &1 &
   &1 &
   &1 & A$_1$ \\
    1 & 
   &3 & 
   &0 &
$-$&1 &
   &1 &
$-$&1 & F$_1$  \\
    2 & 
   &5 & 
$-$&1 &
   &1 &
$-$&1 &
   &1 & $\text{E}\oplus \text{F}_2$ \\
    3 & 
   &7 & 
   &1 &
$-$&1 &
$-$&1 &
$-$&1 & $\text{A}_2\oplus\text{F}_1\oplus\text{F}_2$ \\
    4 & 
   &9 & 
   &0 &
   &1 &
   &1 &
   &1 & $\text{A}_1\oplus\text{E}\oplus\text{F}_1\oplus\text{F}_2$ \\
  5 & 
  &11 & 
$-$&1 &
$-$&1 &
   &1 &
$-$&1 & $\text{E}\oplus 2\text{F}_1\oplus \text{F}_2$ \\
  \cline{1-12} \\[-0.4cm]
  \cline{1-12} \\[-0.4cm]           
  \end{tabular}  
}
\end{table}

\begin{table}
  \caption{%
    Characters and irrep decomposition
    of the rotational functions of water
    in the $C_\text{2v}$(M) molecular symmetry group.
    \label{tab:charwat}
  }
\scalebox{0.9}{%
  \begin{tabular}{@{}l r@{}l r@{}l r@{}l r@{}l r@{}l | r@{}l r@{}l r@{}l r@{}l r@{}l @{}}
  \cline{1-9} \\[-0.4cm]
  \cline{1-9} \\[-0.4cm]
    & 
    \multicolumn{2}{c}{$E$} & 
    \multicolumn{2}{c}{(ab)} & 
    \multicolumn{2}{c}{$E^\ast$} &
    \multicolumn{2}{c}{$[(\mr{ab})]^\ast$} \\    
     & 
    \multicolumn{2}{c}{1} & 
    \multicolumn{2}{c}{1} & 
    \multicolumn{2}{c}{1} & 
    \multicolumn{2}{c}{1} \\        
  \cline{1-9} \\[-0.4cm]
    $\jwat_{\ka,\kc}$
    & 
    \multicolumn{2}{c}{1} & 
    \multicolumn{2}{c}{$(-1)^{\ka+\kc}$} & 
    \multicolumn{2}{c}{$(-1)^{\kc}$} &
    \multicolumn{2}{c}{$(-1)^{\ka}$} \\    
  \cline{1-9} \\[-0.4cm]
  \cline{1-9} \\[-0.4cm]           
  \end{tabular}
}  
\end{table}

%
%
\begin{table}
\caption{%
  Characters and irrep decomposition of the 
  $[[\jmet,\jwat_{\ka\kc}]_j,\Lambda]_J$
  coupled-rotor functions which dominate 
  the ZPV(GM) splitting manifold with $J=0$ 
  in the $G_{48}$ group
  (see also Figures~\ref{fig:rovibgp} and \ref{fig:rovibgm}).
  \label{tab:crsymmj0}
}
\scalebox{0.8}{%
  \begin{tabular}{@{}l r@{\ \ }l r@{}l r@{}l r@{}l r@{}l  r@{}l r@{}l r@{}l r@{}l r@{}l c @{}}
  \cline{1-22} \\[-0.4cm]
  \cline{1-22} \\[-0.4cm]
    $\Gamma$ & 
    \multicolumn{2}{c}{\small $E$} & 
    \multicolumn{2}{c}{\small (123)} & 
    \multicolumn{2}{c}{\small (14)(23)} & 
    \multicolumn{2}{c}{\small [(1423)(ab)]$^\ast$} & 
    \multicolumn{2}{c}{\small [(23)(ab)]$^\ast$} & 
    \multicolumn{2}{c}{\small (ab)} & 
    \multicolumn{2}{c}{\small (123)(ab)} & 
    \multicolumn{2}{c}{\small (14)(23)(ab)} & 
    \multicolumn{2}{c}{\small [(1423)]$^\ast$} & 
    \multicolumn{2}{c}{\small [(23)]$^\ast$} & 
    Irreps \\    
    \multicolumn{1}{r}{} & 
    \multicolumn{2}{c}{1} & 
    \multicolumn{2}{c}{8} & 
    \multicolumn{2}{c}{3} & 
    \multicolumn{2}{c}{6} & 
    \multicolumn{2}{c}{6} & 
    \multicolumn{2}{c}{1} & 
    \multicolumn{2}{c}{8} & 
    \multicolumn{2}{c}{3} & 
    \multicolumn{2}{c}{6} & 
    \multicolumn{2}{c}{6} & \\        
  \cline{1-22} \\[-0.4cm]
  %
 $[[0,0_{00}]_0,0]_{0}$ & &1 & &1 & &1 & &1 & &1 & &1 & &1 & &1 & &1 & &1 & $\Aonep$ \\
 $[[0,1_{11}]_1,1]_{0}$ & &1 & &1 & &1 & &1 & &1 & &1 & &1 & &1 & &1 & &1 & $\Aonep$ \\
 $[[1,0_{00}]_1,1]_{0}$ & &3 & &0 & $-$&1 & $-$&1 & &1 & &3 & &0 & $-$&1 & $-$&1 & &1 & $\Ftwop$\\
 $[[1,1_{11}]_0,0]_{0}$ & &3 & &0 & $-$&1 & $-$&1 & &1 & &3 & &0 & $-$&1 & $-$&1 & &1 & $\Ftwop$ \\
 $[[1,1_{11}]_1,1]_{0}$ & &3 & &0 & $-$&1 & &1 & $-$&1 & &3 & &0 & $-$&1 & &1 & $-$&1 & $\Fonep$ \\
 $[[1,1_{11}]_2,2]_{0}$ & &3 & &0 & $-$&1 & $-$&1 & &1 & &3 & &0 & $-$&1 & $-$&1 & &1 & $\Ftwop$ \\
 $[[2,0_{00}]_2,2]_{0}$ & &5 & $-$&1 & &1 & $-$&1 & &1 & &5 & $-$&1 & &1 & $-$&1 & &1 & $\Ep\oplus\Ftwop$ \\
 $[[2,1_{11}]_1,1]_{0}$ & &5 & $-$&1 & &1 & $-$&1 & &1 & &5 & $-$&1 & &1 & $-$&1 & &1 & $\Ep\oplus\Ftwop$ \\
 $[[2,1_{11}]_2,2]_{0}$ & &5 & $-$&1 & &1 & &1 & $-$&1 & &5 & $-$&1 & &1 & &1 & $-$&1 & $\Ep\oplus\Fonep$ \\
 $[[2,1_{11}]_3,3]_{0}$ & &5 & $-$&1 & &1 & $-$&1 & &1 & &5 & $-$&1 & &1 & $-$&1 & &1 &  $\Ep\oplus\Ftwop$ \\
  \cline{1-22} \\[-0.4cm]
  %
 $[[0,1_{01}]_1,1]_{0}$ & &1 & &1 & &1 & $-$&1 & $-$&1 & $-$&1 & $-$&1 & $-$&1 & &1 & &1 & $\Atwom$ \\
 $[[0,1_{10}]_1,1]_{0}$ & &1 & &1 & &1 & &1 & &1 & $-$&1 & $-$&1 & $-$&1 & $-$&1 & $-$&1 & $\Aonem$ \\
 $[[1,1_{01}]_0,0]_{0}$ & &3 & &0 & $-$&1 & &1 & $-$&1 & $-$&3 & &0 & &1 & $-$&1 & &1 & $\Fonem$ \\
 $[[1,1_{01}]_1,1]_{0}$ & &3 & &0 & $-$&1 & $-$&1 & &1 & $-$&3 & &0 & &1 & &1 & $-$&1 & $\Ftwom$ \\
 $[[1,1_{01}]_2,2]_{0}$ & &3 & &0 & $-$&1 & &1 & $-$&1 & $-$&3 & &0 & &1 & $-$&1 & &1 & $\Fonem$ \\
 $[[1,1_{10}]_0,0]_{0}$ & &3 & &0 & $-$&1 & $-$&1 & &1 & $-$&3 & &0 & &1 & &1 & $-$&1 & $\Ftwom$ \\
 $[[1,1_{10}]_1,1]_{0}$ & &3 & &0 & $-$&1 & &1 & $-$&1 & $-$&3 & &0 & &1 & $-$&1 & &1 & $\Fonem$ \\
 $[[1,1_{10}]_2,2]_{0}$ & &3 & &0 & $-$&1 & $-$&1 & &1 & $-$&3 & &0 & &1 & &1 & $-$&1 & $\Ftwom$ \\
 $[[2,1_{01}]_1,1]_{0}$ & &5 & $-$&1 & &1 & &1 & $-$&1 & $-$&5 & &1 & $-$&1 & $-$&1 & &1 & $\Em\oplus\Fonem$ \\
 $[[2,1_{01}]_2,2]_{0}$ & &5 & $-$&1 & &1 & $-$&1 & &1 & $-$&5 & &1 & $-$&1 & &1 & $-$&1 & $\Em\oplus\Ftwom$ \\
 $[[2,1_{01}]_3,3]_{0}$ & &5 & $-$&1 & &1 & &1 & $-$&1 & $-$&5 & &1 & $-$&1 & $-$&1 & &1 & $\Em\oplus\Fonem$ \\
 $[[2,1_{10}]_1,1]_{0}$ & &5 & $-$&1 & &1 & $-$&1 & &1 & $-$&5 & &1 & $-$&1 & &1 & $-$&1 & $\Em\oplus\Ftwom$ \\
 $[[2,1_{10}]_2,2]_{0}$ & &5 & $-$&1 & &1 & &1 & $-$&1 & $-$&5 & &1 & $-$&1 & $-$&1 & &1 & $\Em\oplus\Fonem$ \\
 $[[2,1_{10}]_3,3]_{0}$ & &5 & $-$&1 & &1 & $-$&1 & &1 & $-$&5 & &1 & $-$&1 & &1 & $-$&1 & $\Em\oplus\Ftwom$ \\
  \cline{1-22} \\[-0.4cm]
  \cline{1-22} \\[-0.4cm]           
  \end{tabular}
}  
\end{table}

\clearpage
\section{Numerical results and discussion}
\subsection{Rovibrational energy-level pattern of the methane-water dimers}
\noindent%
In this section, the computed rovibrational states with $J=0,1$ and 2
are presented for the four studied C$X_4\cdot Y_2\text{O}$ ($X,Y=$~H, D) dimers focusing on the zero-point vibrational splitting of the global minimum, ZPV(GM).
The computed energy levels are provided in Tables~S1--S12 of the Supplementary Information. The energy-level differences can be directly compared with high-resolution spectroscopic measurements, when such measurements will be become available for the deuterated isotopologues (the electric dipole selection rules and the spin-statistical weights of HH are discussed in detail in Ref.~\cite{SaCsAlWaMa16}). 

The zero-point vibrational state, or any other vibrational state symmetric with respect to the symmetry plane of the equilibrium structure, is split into the following direct sum, including 24 states \cite{SaCsAlWaMa16}:
\begin{align}
  \mathit{\Gamma}_{\mr{A}'} 
  &= 
  \Aonep \oplus \Ep \oplus \Fonep \oplus 2\Ftwop \oplus 
  \Atwom \oplus \Em \oplus 2\Fonem \oplus \Ftwom 
  \label{eq:gamsym} 
\end{align}
in the $G_{48}$ molecular symmetry group.
The $+$ and $-$ superscripts refer to the symmetry and anti-symmetry 
of the spatial wave function with respect to the exchange of the 
two protons(deuterons) of H$_2$O(D$_2$O) in agreement with the labeling of Ref.~\cite{DoSa94}.

As to the vibrational states, 
an irrep label can be assigned to the computed vibrational states by visual inspection of one- and two-dimensional cuts of the wave
function, as it was done in Ref.~\cite{SaCsAlWaMa16} (for multidimensional representations the trace of the representation matrix had to be taken). In the present work, the CRD decomposition and the symmetry rules derived in the previous section made it possible to carry out the symmetry assignment in an almost automated fashion, which is particularly useful for the assignment of the multi-dimensional irreps. 

\begin{figure}
  \includegraphics[width=15cm]{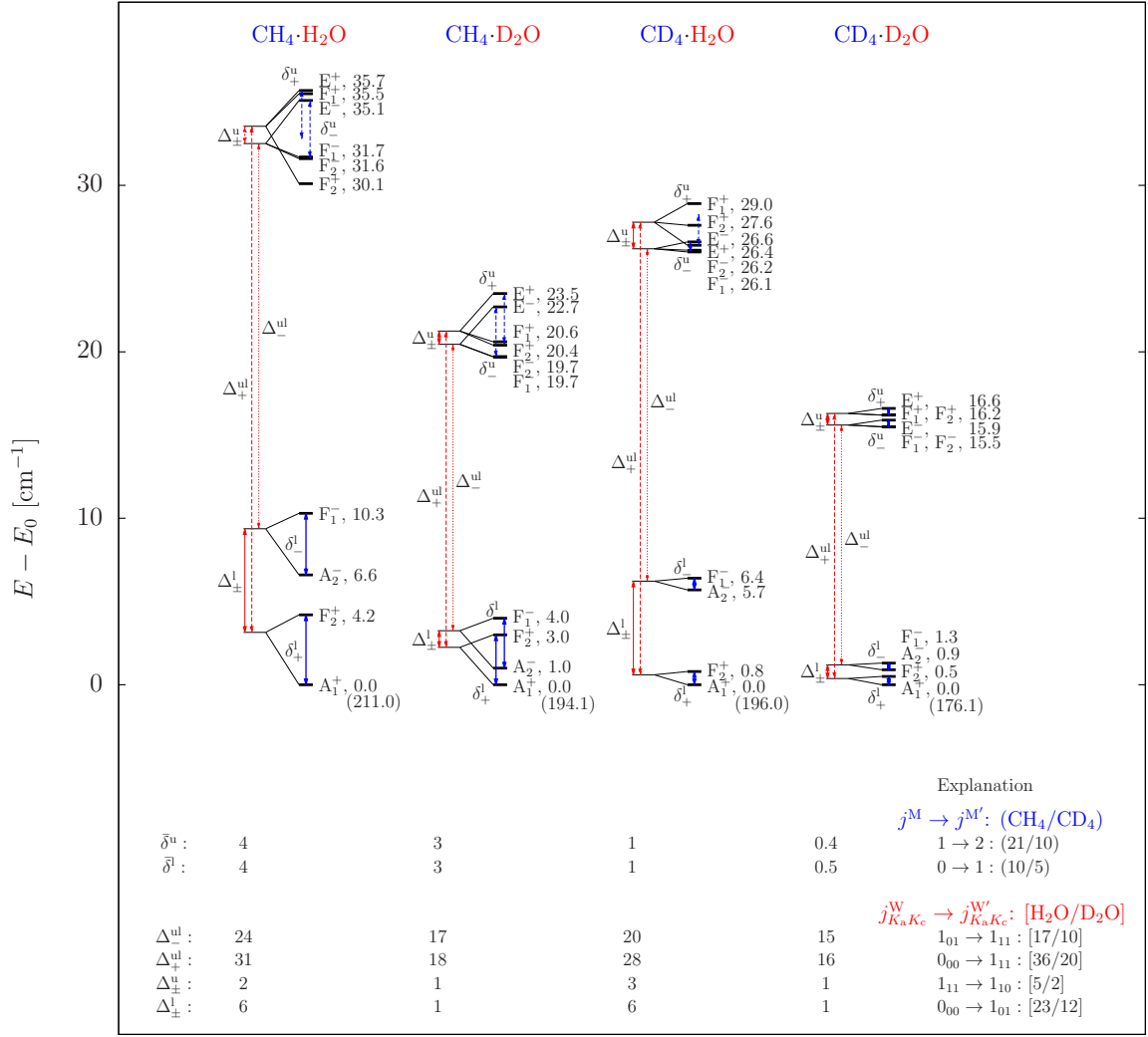}
  \caption{%
    Zero-point vibrational splitting energies,  in \cm, assigned to 
    the global minimum of the four isotopologues studied.
    The splittings indicated with double-headed arrows are color coded with
    blue and red referring to the methane and water monomers, respectively.
    \label{fig:zpve}
  }
\end{figure}

\begin{figure}  
\includegraphics[width=20cm,angle=-90]{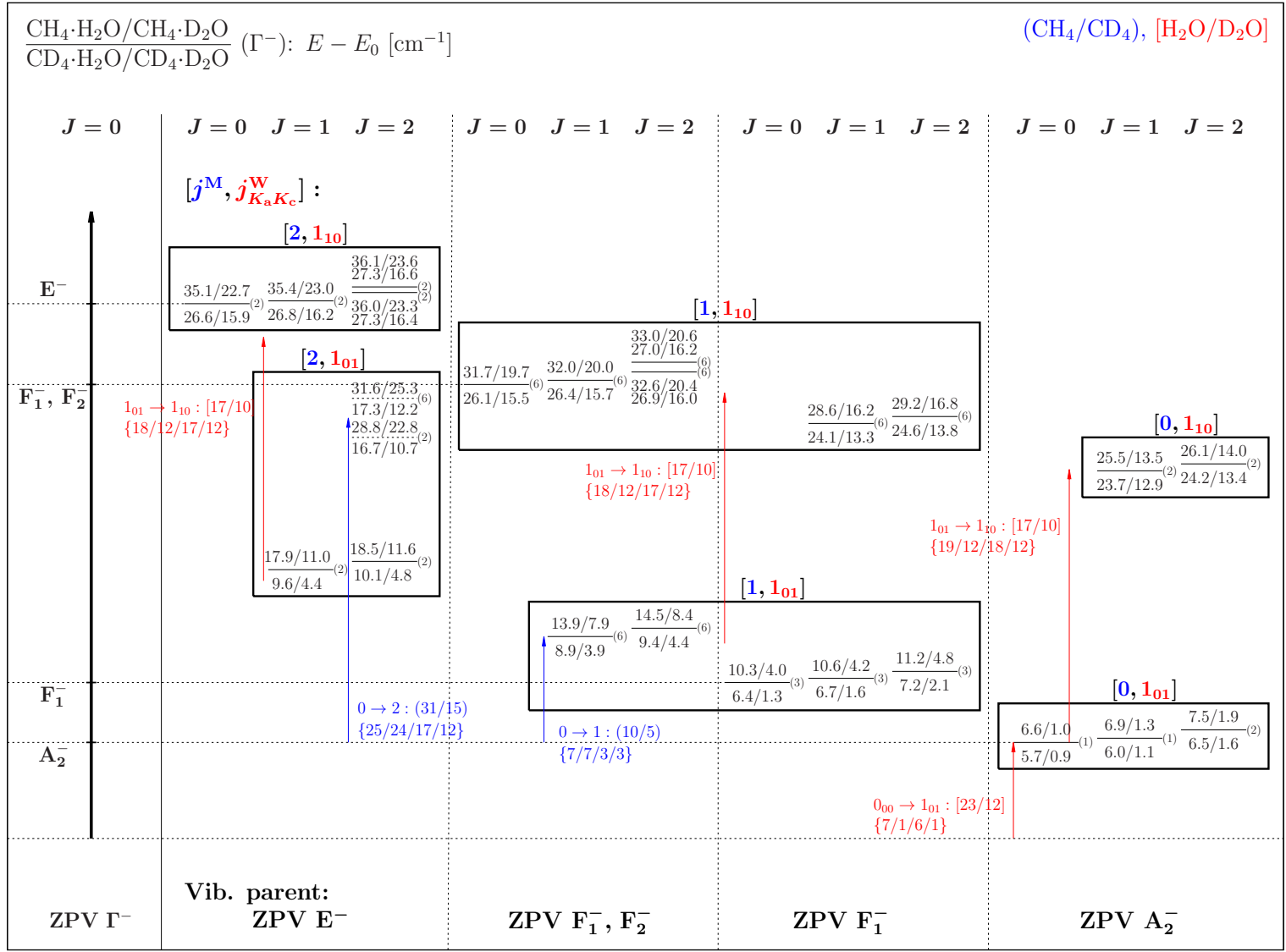} 
  \caption{%
    Summary of $J=0,1$, and 2 rovibrational states of the
    methane-water dimers with $\Gamma^-$ symmetry.
    Levels shown with dashed lines in the ZPV, $\Em$ column 
    (for all studied isotopologues) correspond to the 
    $(2,1_{01})$ monomer rotor states but they cannot be 
    assigned to a single, dominant vibrational parent state.
    \label{fig:rovibgm}
  }
\end{figure}

This automated assignment scheme facilitated the identification of
a mistake in our earlier assignment of the 1 and 2 subindices of the triply degenerate irreps of HH \cite{SaCsAlWaMa16}. According to the corrected assignment of the ZPV(GM) splitting ($J=0$), there is an $\Ftwop$ and an $\Fonem$ state in the lower part and there is an ($\Fonep$, $\Ftwop$) pair as well as an 
($\Fonem$, $\Ftwom$) pair of states in the upper part of the splitting with $J=0$ (see Figures~\ref{fig:zpve}--\ref{fig:rovibgm}).

To characterize the rovibrational states (Figures~\ref{fig:rovibgp} and \ref{fig:rovibgm}), we identified the parent vibrational state(s) of each rovibrational state using the rigid-rotor decomposition (RRD) scheme \cite{NMD10}. When the identification of a single parent vibrational state 
is possible, the rovibrational state can be imagined as some rotational 
excitation of the parent vibrational state.  

The numerical values of the energy 
levels and the splittings are very different for the four isotopologues studied, but the overall rovibrational pattern and assignment is very similar (recall that the four systems are described on the same PES but have different monomer rotational constants): 
both the vibrational and the rovibrational energy-level pattern is characterized by a lower and an upper part split into further levels (see Figs.~\ref{fig:zpve}--\ref{fig:rovibgm}). 
As it can be expected, the energy values get smaller when the monomers are replaced with their heavier (deuterated) versions.

More interestingly, all studied methane-water dimers have an unusual 
feature: there are rotational ``excitations'' which are lower in energy than the energy of their vibrational parent state; thus, 
these rovibrational states appear as a rotational excitation with a 
negative transition energy. These reversed rovibrational sequences were 
present in the experimental data reported for CH$_4\cdot$H$_2$O \cite{DoSa94} 
and were also reproduced in the computations of Ref.~\cite{SaCsAlWaMa16}. 
Reversed rovibrational sequences have also been identified computationally in other systems, for example, in H$_5^+$ \cite{FaSaCs14} as well as in the CH$_5^+$ molecular ion \cite{WoMa15,ScJeSc15,WaCa16}. These reversed sequences observed in the computations for 
H$_5^+$ were explained by a 2-dimensional torsional-rotational model \cite{FaSaCs14}. Furthermore, Refs.~\cite{ScJeSc15,ScJeSc16,ScJeSc17} explained this unusual behavior by treating some large-amplitude motions (LAM) on the same footing as the rotations by going beyond the SO(3) group for these LAM-rotational degrees of freedom.

\begin{figure}
  \includegraphics[width=7.8cm]{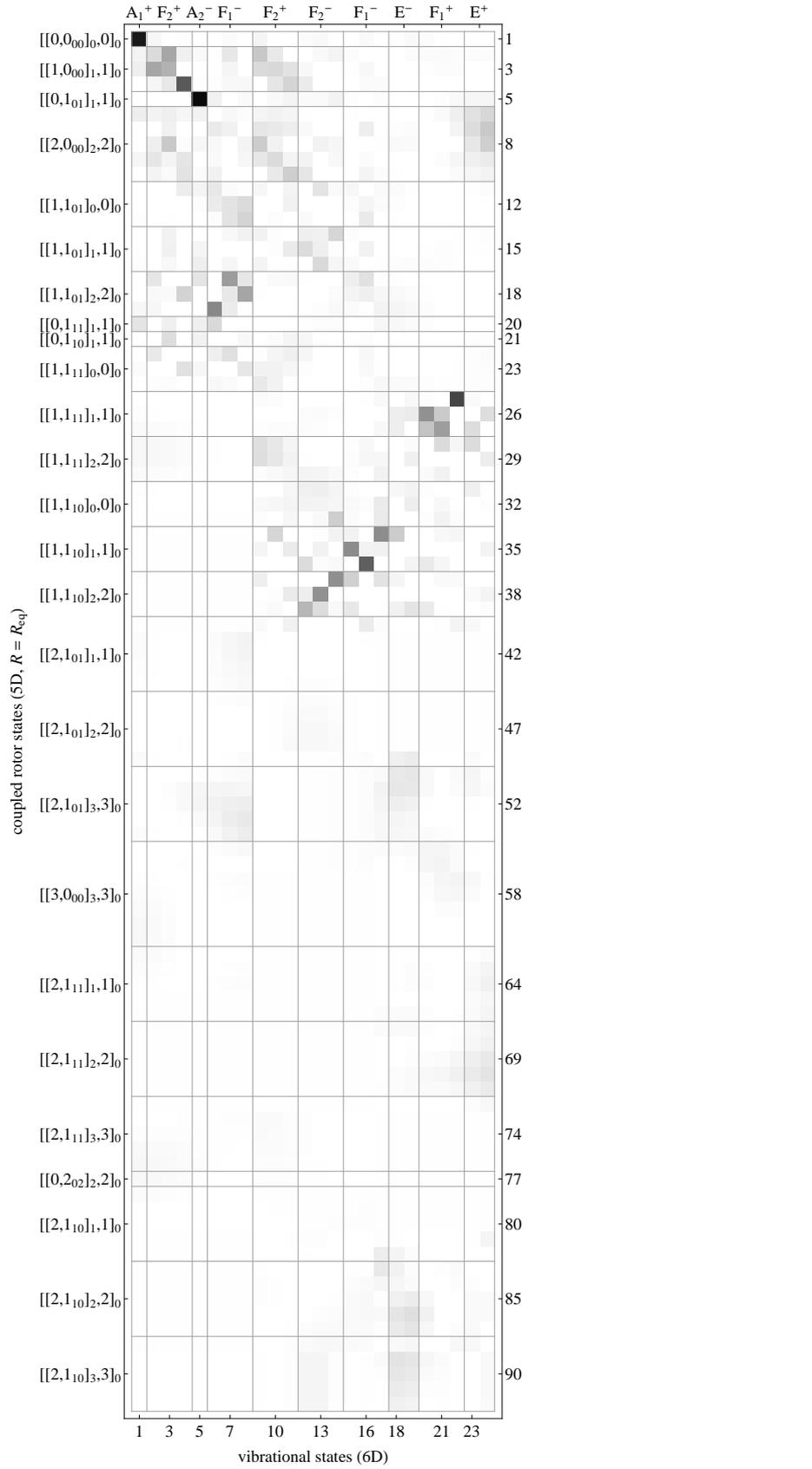} 
  \caption{%
    Coupled-rotor decomposition (CRD) table for \metwata\
    with $J=0$: overlap (Eq.~\ref{eq:defcrd}) of the 24 lowest-energy vibrational states, which belong to the ZPV(GM) splitting manifold, with the 92 lowest-energy coupled-rotor states. 
    \label{fig:exCRDJ0}
  }
\end{figure}

In the present work, we take a simple strategy to better understand the origin of the splitting pattern and these formally negative-energy rotational excitations: besides the rigid-rotor model, which underlies the assignment and the observation of these reversed sequences, we use the coupled-rotor system as another meaningful model for weakly interacting dimers. The computed coupled-rotor decomposition (CRD, introduced in Section~\ref{ch:assign}) tables provide detailed information about the monomers' rotational states and their relative rotation within the rotating dimer.
In most of this work, we focus on the assignment of monomer-rotor labels, $[\jmet,\jwat_{\ka,\kc}]$, because the monomer excitations are about an order of magnitude more energetic than the coupling (in particular, the water's rotational constants are much larger than the dimer's rotational constants).
For this purpose, it is sufficient to use CRD tables computed with a large intermonomer separation, the one used is $R_{100}=100$~bohr. In this case the CR energy is the simple sum of the monomer rotor energies (Tables~S13--S19). 
To better understand the fine details of the dimer's internal dynamics, we also assigned $j$ and $\Lambda$ for a few states using CR functions obtained with monomers fixed at the complex's equilibrium distance, $R=R_{\text{eq}}=6.5$~bohr. For this $R$ value the $j$-splitting of the CR energies is small but larger than the convergence of our variational results, so the assignment of the $j$ and $\Lambda$ labels (Section~\ref{ch:crsymm}) is possible. Figure~\ref{fig:exCRDJ0}  presents an example of a complete CR assignment.

The results of the symmetry, RR and CR assignments of the computed rovibrational states for all four isotopologues are collected in Figures~\ref{fig:rovibgp} and \ref{fig:rovibgm} 
(detailed results are provided in Tables~S1--S12 of the \som).
As a general and important conclusion which becomes transparent from the CRD assignment, the characteristic upper-lower separation of the rovibrational ZPV(GM) manifold corresponds to the 
$1_{11}\leftarrow 0_{00}$ and the $1_{10}\leftarrow 1_{01}$ rotational excitation of the water monomer for states of $\Gamma^+$ and $\Gamma^-$ symmetry (para and ortho for HH and DH), respectively. The sub-splittings of both the lower and the upper parts are related to the rotational excitation of the methane monomer with one and two quanta and to the excitation of the relative rotation of the two monomers.

\subsection{Reversed rovibrational sequences}
\noindent%
The RRD assignment of the rovibrational states to their vibrational parents highlighted that certain rotational excitations have ``negative'' transition energies. These reversed rovibrational sequences are observed for all four isotopologues and they appear within the vibrational bands of certain vibrational symmetries.

The CRD tables show that these negative rotational excitation energies correspond to a $J+1\leftarrow J$ transition in which the water molecule looses rotational energy---$0_{00}\leftarrow 1_{11}$ and $1_{01}\leftarrow 1_{10}$ for $\Gamma^+$ and $\Gamma^-$, respectively--- but due to a different internal coupling of the methane and the water rotors the total rotational quantum number increases. This observation is certainly useful and provides an intuitive picture about this phenomenon but it remains a question to be answered why the lower-energy $J=0$ state is ``missing'' (and is replaced by its higher-energy analogue) in several vibrational bands ($\Ep$, $\Fonep$, $\Ftwop$, $\Em$, $\Fonem$, $\Fonem$).
Is that coupling symmetry forbidden or does this pattern result from a particular mixing of the CR states?

\begin{figure}
  \includegraphics[width=20cm,angle=-90]{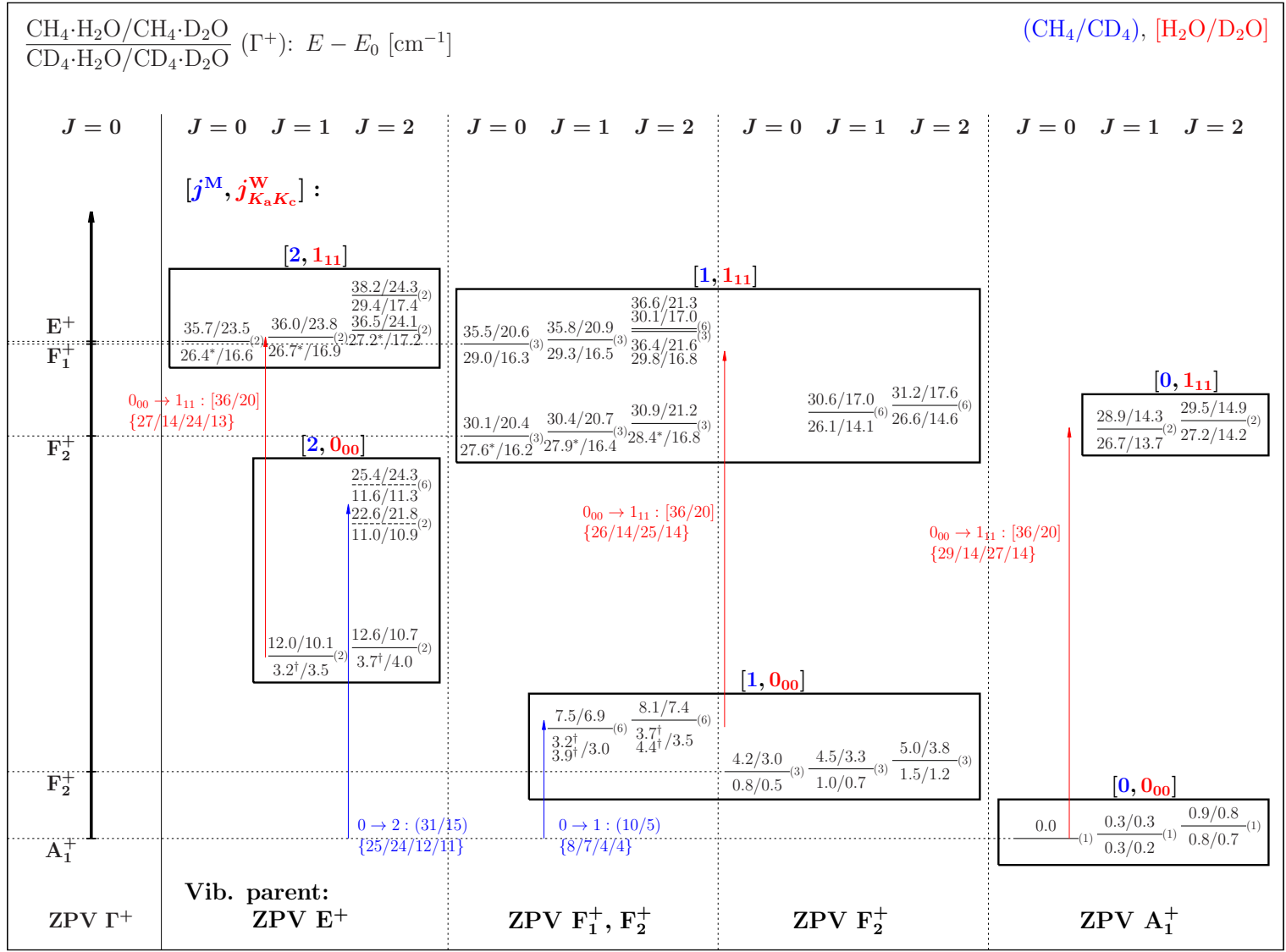}
  \caption{%
    Summary of $J=0,1$, and 2 rovibrational states of the methane-water 
    dimers with $\Gamma^+$ symmetry.
    Levels shown with dashed lines in the ZPV, $\Ep$ 
    column (for all studied isotopologues) 
    correspond to the $(2,0_{00})$ monomer rotor states 
    of the ZPV rovibrational band but cannot be 
    assigned to a single, dominant vibrational state.
    States of the \metwatc\ isotopologue are somewhat outliers from 
    the general rovibrational pattern.
    The states labelled with $^\ast$ are characterized by 
    a strong mixing of the
    $(1,0_{00})$, $(2,0_{00})$, $(1,1_{11})$, and $(2,1_{11})$ 
    monomer rotor states.
    The states labelled with $^\dagger$ exhibit a 
    strong mixing with the $\Ep$
    vibrational parent levels in the RRD.
    \label{fig:rovibgp}
  }
\end{figure}

In the case of the  $\Fonep$-symmetry ZPV(GM) vibration (see Fig.~\ref{fig:rovibgp})
the negative-energy rotational transition 
is dictated by symmetry rules (Tables~\ref{tab:crsymmj0}, S23, and S25). The $[[1,1_{11}]_1,1]_0$ and $[[2,1_{11}]_2,2]_0$ states are the lowest-energy CR states of $\Fonep$ symmetry with $J=0$. They correspond to rotationally excited water, so they are more energetic and appear in the ``upper'' part of the splitting pattern. At the same time, the $J=1$ and 2 rotational excitations of this vibrational state can have contributions from CR states with water in its lowest rotational state, $0_{00}$, which lowers the energy of these rovibrational states and makes them appear in the lower part of the splitting. As a result of these symmetry properties,  the rotational excitation of the $\Fonep$ vibrational state shows up as a ``negative-energy'' transition.

Unfortunately, the explanation of the reversed rovibrational sequences corresponding to other vibrational states is less straightforward when based on simple symmetry-related arguments (Table~\ref{tab:crsymmj0} and Tables~S23--S26). For example, the $\Ep$ vibrational band has only a ``higher-energy'' vibrational state but there are both ``lower-energy'', $[2,0_{00}]$, and ``higher-energy'', $[2,1_{11}]$, CR functions of $\Ep$ symmetry with $J=0$. In fact, we observe in the numerically computed CRD tables that the lower-energy $[2,0_{00}]$ CR functions contribute to many states but they do not dominate any of the states. Namely, the $J=0$ $\Ep$ vibration is assigned to the higher-energy $[2,1_{11}]$ CR function with some contribution also from $[2,0_{00}]$. The HH isotopologue, whose assignment slightly differs from the other three isotopologues, has about equal contributions from both $[2,1_{11}]$ and $[2,0_{00}]$. 
Similar behavior is observed for the $\Ftwop$, $\Em$, $\Fonem$, and $\Ftwom$ vibrational bands of the ZPV(GM). The reversed energy ordering results from a particular mixing of CR states with lower-energy and higher-energy water rather than being due to strict symmetry rules. We also note that for the $\Ftwop$ and $\Fonem$ states of the ZPV(GM) there are both reversed and normal sequences. 

In general, the computed CRD tables show that the higher-energy rovibrational states are dominated by higher-energy CR functions with the $1_{11}$ and $1_{10}$ higher-energy rotational states of water in $\Gamma^+$ and $\Gamma^-$, respectively. At the same time, the lower-energy states are dominated by lower-energy CR functions with the $0_{00}$ and $1_{01}$ excited rotational states of water in $\Gamma^+$ and $\Gamma^-$, respectively.

\clearpage

\clearpage

\clearpage
\section{Summary and outlook}
\noindent%
Rovibrational states of four isotopologues of the methane-water dimer have been computed using a six-dimensional intermolecular vibrational model utilizing the GENIUSH code \cite{MaCzCs09,FaMaCs11} and the QCHB15 potential energy surface \cite{QuCoHoBo15}. The computed rovibrational states (with $J=0,1,$ and 2 rotational quantum numbers) were assigned to their parent vibrational state(s) using the rigid-rotor decomposition (RRD) scheme. This assignment highlighted in all four dimers,  \metwata, \metwatb, \metwatc, and \metwatd, the existence of a reversed rovibrational energy ordering, \emph{i.e.,} formally ``negative'' rotational excitation energies. 
 
In order to better understand the complex rovibrational dynamics of this fluxional, weakly bound complex, we developed the coupled-rotor decomposition (CRD) scheme.
The CRD scheme is a general approach to determine the monomers' rotational states and their relative rotation within a rotating dimer.

For the example of the methane-water dimer the symmetry properties of the coupled-rotor functions were derived
employing the $G_{48}$ molecular symmetry (MS) group.
Together with the numerically computed CRD tables this symmetry classification opens the route to an automated symmetry assignment of the rovibrational states of the dimer, a particularly useful approach for higher excited, multiply degenerate states.

As to the interpretation of the higher-energy states of the methane-water dimer, we can think of a number of challenging, open questions. Is it possible to identify the zero-point splitting manifold of the secondary minimum, ZPV(SM), similarly to the ZPV(GM) studied in the present work? Is the full ZPV(SM) bound or does it spread beyond the first dissociation threshold (which is only $D_0(\text{HH})=152.030$~\cm\ for \metwata\ on the QCHB15 PES \cite{QuCoHoBo15}). Furthermore, by thinking in terms of the coupled-rotors picture, we may imagine energetic monomer rotational states which bring the system above the dissociation energy. Since the intermolecular angular degrees of freedom are only weakly coupled to the intermonomer stretching (the dissociation coordinate), we may expect long-lived quasi-bound states in which the energy is stored in spinning monomers. 
These questions highlight the extremely rich internal-rotational dynamics of this weakly bound dimer, which is worth of further experimental and computational inquiries.
Seemingly all the theoretical tools are at hand to help understanding high-resolution far-infrared spectra of these complexes.

\vspace{1cm}
\noindent %
\textbf{Acknowledgment} \\
J.S. received support from the ÚNKP-16-3 New
National Excellence Program of the Ministry of Human Capacities of Hungary. 
The work performed by A.G.C. received support from NKFIH
(grant no. K119658).
E.M. acknowledges financial support from a PROMYS Grant (no. IZ11Z0\_166525)  
of the Swiss National Science Foundation.

\clearpage

\end{document}


\title{%
{\sc Supplementary Information to} \\
Rovibrational quantum dynamical computations for deuterated isotopologues of the methane-water dimer%
}
\author{J\'anos Sarka}
\affiliation{Institute of Chemistry, E\"otv\"os Lor\'and University, P\'azm\'any P\'eter s\'et\'any 1/A, Budapest, H-1117, Hungary}
\affiliation{MTA-ELTE Complex Chemical Systems Research Group, P.O. Box 32, H-1518 Budapest 112, Hungary}

\author{Attila G. Cs\'asz\'ar}
\affiliation{Institute of Chemistry, E\"otv\"os Lor\'and University, P\'azm\'any P\'eter s\'et\'any 1/A, Budapest, H-1117, Hungary}
\affiliation{MTA-ELTE Complex Chemical Systems Research Group, P.O. Box 32, H-1518 Budapest 112, Hungary}

\author{Edit M\'atyus}
\email{matyus@chem.elte.hu}
\affiliation{Institute of Chemistry, E\"otv\"os Lor\'and University, P\'azm\'any P\'eter s\'et\'any 1/A, Budapest, H-1117, Hungary}

\date{\today}

\maketitle 

\clearpage
\setcounter{section}{0}
\renewcommand{\thesection}{S\arabic{section}}
\setcounter{subsection}{0}
\renewcommand{\thesubsection}{S\arabic{section}.\arabic{subsection}}

\setcounter{equation}{0}
\renewcommand{\theequation}{S\arabic{equation}}

\setcounter{table}{0}
\renewcommand{\thetable}{S\arabic{table}}

\setcounter{figure}{0}
\renewcommand{\thefigure}{S\arabic{figure}}

\section{Tables}


\begin{table}[ht!]
\caption{%
  Vibrational states of \metwata\ (HH) with $J=0$: 
  energy levels, irrep labels of the $G_{48}$ molecular symmetry group, and coupled-rotor decomposition.
  \label{tab:hhj0}
}
\begin{tabular}{@{}l@{\ \ }c@{\ \ }c@{\ \ }l@{\ \ }l@{}}
\hline\hline
\multicolumn{1}{l}{Label} & \multicolumn{1}{c}{$E$ [\cm]} & \multicolumn{1}{c}{$\Gamma^{\pm}$} & \multicolumn{2}{c}{Coupled-rotor states ($J=0$)} \\ \cline{4-5}\\[-0.4cm]
 & & & \multicolumn{1}{l}{dominant} & \multicolumn{1}{l}{minor} \\
\hline
     HH--J0.1 & 219.6 & $\aep$ & $[0,0_{00}]$ & $[3,0_{00}]$ \\
  HH--J0.2--4 & 4.2   & $\fkp$ & $[1,0_{00}]$ & $[2,0_{00}]$ \\
     HH--J0.5 & 6.6   & $\akm$ & $[0,1_{01}]$ & ~ \\
  HH--J0.6--8 & 10.3  & $\fem$ & $[1,1_{01}]$ & $[2,1_{01}]$ \\
 HH--J0.9--11 & 30.1  & $\fkp$ & $[1,1_{11}]$ & $[1,0_{00}]$ \& $[2,0_{00}]$ \\
HH--J0.12--14 & 31.6  & $\fkm$ & $[1,1_{10}]$ & $[1,1_{01}]$ \& $[2,1_{10}]$ \\
HH--J0.15--17 & 31.7  & $\fem$ & $[1,1_{10}]$ & $[1,1_{01}]$ \& $[2,1_{10}]$ \\
HH--J0.18--19 & 35.1  & $\em$  & $[2,1_{10}]$ & $[2,1_{01}]$\\
HH--J0.20--22 & 35.5  & $\fep$ & $[1,1_{11}]$ & ~                    \\
HH--J0.23--24 & 35.7  & $\ep$  & $[2,0_{00}]$ \& $[2,1_{11}]$ & \\
    HH--J0.25 & 36.1 & $\aep$(SM) & $[0,0_{00}]$ & $[0,1_{11}]$ \& $[3,0_{00}]$ \\

\hline\hline
\end{tabular}
\end{table}

\begin{table}[ht!]
\caption{%
  Vibrational states of \metwatb\ (HD) with $J=0$. 
  See also caption to Table~\ref{tab:hhj0}. 
  \label{tab:hdj0}
}
\begin{tabular}{@{}l@{\ \ }c@{\ \ }c@{\ \ }l@{\ \ }l@{}}
\hline\hline
\multicolumn{1}{l}{Label} & \multicolumn{1}{c}{$E$ [\cm]} & \multicolumn{1}{c}{$\Gamma^{\pm}$} & \multicolumn{2}{c}{Coupled-rotor states ($J=0$)} \\ \cline{4-5}\\[-0.4cm]
 & & & \multicolumn{1}{l}{dominant} & \multicolumn{1}{l}{minor} \\
\hline
     HD--J0.1 & 202.8 & $\aep$ & $[0,0_{00}]$ & $[0,2_{02}]$ \& $[3,0_{00}]$ \\
     HD--J0.2 & 1.0   & $\akm$ & $[0,1_{01}]$ & $[3,1_{01}]$ \\
  HD--J0.3--5 & 3.0   & $\fkp$ & $[1,0_{00}]$ & $[1,2_{02}]$ \& $[2,0_{00}]$ \\
  HD--J0.6--8 & 4.0   & $\fem$ & $[1,1_{01}]$ & $[2,1_{01}]$ \\
 HD--J0.9--11 & 19.7  & $\fem$ & $[1,1_{10}]$ & ~ \\
HD--J0.12--14 & 19.7  & $\fkm$ & $[1,1_{10}]$ & ~ \\
HD--J0.15--17 & 20.4  & $\fkp$ & $[1,1_{11}]$ & $[1,2_{11}]$ \\
HD--J0.18--20 & 20.6  & $\fep$ & $[1,1_{11}]$ & $[1,2_{11}]$ \\
HD--J0.21--22 & 22.7  & $\em$  & $[2,1_{10}]$ & ~            \\
HD--J0.23--24 & 23.5  & $\ep$  & $[2,1_{11}]$ & ~            \\
\hline\hline
\end{tabular}
\end{table}

\begin{table}[ht]
\caption{%
  Vibrational states of \metwatc\ (DH) with $J=0$.
  See also caption to Table~\ref{tab:hhj0}. 
  \label{tab:dhj0}
}
\begin{tabular}{@{}l@{\ \ }c@{\ \ }c@{\ \ }l@{\ \ }l@{}}
\hline\hline
\multicolumn{1}{l}{Label} & \multicolumn{1}{c}{$E$ [\cm]} & \multicolumn{1}{c}{$\Gamma^{\pm}$} & \multicolumn{2}{c}{Coupled-rotor states ($J=0$)} \\ \cline{4-5}\\[-0.4cm]
 & & & \multicolumn{1}{l}{dominant} & \multicolumn{1}{l}{minor} \\
\hline
     DH--J0.1 & 204.6 & $\aep$ & $[0,0_{00}]$                 & $[3,0_{00}]$                            \\
  DH--J0.2--4 & 0.8   & $\fkp$ & $[1,0_{00}]$                 & $[2,0_{00}]$                 \\
     DH--J0.5 & 5.7   & $\akm$ & $[0,1_{01}]$                 & $[3,1_{01}]$                 \\
  DH--J0.6--8 & 6.4   & $\fem$ & $[1,1_{01}]$                 & $[2,1_{01}]$                 \\
 DH--J0.9--11 & 26.0  & $\fem$ & $[1,1_{10}]$                 & $[2,1_{10}]$ \& $[3,1_{10}]$ \& $[1,1_{01}]$ \\
DH--J0.12--14 & 26.1  & $\fkm$ & $[1,1_{10}]$                 & $[2,1_{10}]$                 \\
DH--J0.15--16 & 26.4  & $\ep$  & $[1,0_{00}]$ \& $[1,1_{11}]$ & $[2,0_{00}]$ \& $[2,1_{11}]$ \\
DH--J0.17--18 & 26.6  & $\em$  & $[2,1_{10}]$                 & ~                            \\
DH--J0.19--21 & 27.3  & $\fkp$ & $[2,0_{00}]$                 & $[1,1_{11}]$ \& $[2,1_{11}]$ \\
DH--J0.22--24 & 28.9  & $\fep$ & $[1,1_{11}]$                 & $[2,1_{11}]$                 \\
\hline\hline
\end{tabular}
\end{table}

\begin{table}[ht]
\caption{
  Vibrational states of \metwatd\ (DD) with $J=0$.
  See also caption to Table~\ref{tab:hhj0}. 
  \label{tab:ddj0}
}
\begin{tabular}{@{}l@{\ \ }c@{\ \ }c@{\ \ }l@{\ \ }l@{}}
\hline\hline
\multicolumn{1}{l}{Label} & \multicolumn{1}{c}{$E$ [\cm]} & \multicolumn{1}{c}{$\Gamma^{\pm}$} & \multicolumn{2}{c}{Coupled-rotor states ($J=0$)} \\ \cline{4-5}\\[-0.4cm]
 & & & \multicolumn{1}{l}{dominant} & \multicolumn{1}{l}{minor} \\
\hline
     DD--J0.1 & 184.8 & $\aep$ & $[0,0_{00}]$ & $[0,2_{02}]$ \& $[3,0_{00}]$ \\
  DD--J0.2--4 & 0.5   & $\fkp$ & $[1,0_{00}]$ & $[2,0_{00}]$ \& $[1,2_{02}]$ \\
     DD--J0.5 & 0.9   & $\akm$ & $[0,1_{01}]$ & $[3,1_{01}]$              \\
  DD--J0.6--8 & 1.3   & $\fem$ & $[1,1_{01}]$ & $[2,1_{01}]$              \\
 DD--J0.9--11 & 15.5  & $\fem$ & $[1,1_{10}]$ & $[2,1_{10}]$              \\
DD--J0.12--14 & 15.5  & $\fkm$ & $[1,1_{10}]$ & $[2,1_{10}]$              \\
DD--J0.15--16 & 15.9  & $\em$  & $[2,1_{10}]$ & ~                      \\
DD--J0.17--19 & 16.2  & $\fkp$ & $[1,1_{11}]$ & ~                      \\
DD--J0.20--22 & 16.2  & $\fep$ & $[1,1_{11}]$ & $[1,2_{11}]$              \\
DD--J0.23--24 & 16.6  & $\ep$  & $[2,1_{11}]$ & ~                      \\
\hline\hline
\end{tabular}
\end{table}


\begin{table}[ht!]
\caption{%
  Rovibrational states of \metwata\ (HH) with $J=1$: 
  energy levels, irrep labels of the $G_{48}$ group for the the parent vibrational state, 
  and coupled-rotor decomposition.
  \label{tab:hhj1}
}
\begin{tabular}{@{}l@{\ \ }c@{\ \ }l@{\ \ }l@{\ \ }l@{}}
\hline\hline
\multicolumn{1}{l}{Label} & 
\multicolumn{1}{c}{$E$ [\cm]} & 
\multicolumn{1}{l}{Vib. parent} & 
\multicolumn{2}{c}{Coupled-rotor states ($J=1$)} \\ 
\cline{4-5}\\[-0.4cm]
 & & & \multicolumn{1}{l}{dominant} & \multicolumn{1}{l}{minor} \\
\hline
    HH--J1.1      & 0.3   & $\aep$[0.0]       & $[0,0_{00}]$            & ~                    \\
    HH--J1.2--4   & 4.5   & $\fkp$[4.2]       & $[1,0_{00}]$            & $[2,0_{00}]$            \\
    HH--J1.5      & 6.9   & $\akm$[6.6]       & $[0,1_{01}]$            & ~                    \\
    HH--J1.6--11  & 7.5   & $\fkp$[30.1], $\fep$[35.5] & $[1,0_{00}]$            & $[2,0_{00}]$            \\
    HH--J1.12--14 & 10.6  & $\fem$[10.3]      & $[1,1_{01}]$            & $[2,1_{01}]$            \\
    HH--J1.15--16 & 12.0  & $\ep$[35.7]       & $[2,0_{00}]$            & ~                    \\
    HH--J1.17--22 & 13.9  & $\fkm$[31.6], $\fem$[31.7] & $[1,1_{01}]$            & $[2,1_{01}]$\ \&\ $[3,1_{01}]$ \\
    HH--J1.23--24 & 17.9  & $\em$[35.1]       & $[2,1_{01}]$            & ~                    \\
    HH--J1.25--26 & 25.5  & $\akm$[6.6]       & $[0,1_{10}]$            & $[2,1_{01}]$            \\
    HH--J1.27--32 & 28.6  & $\fem$[10.3]      & $[1,1_{10}]$            & $[1,1_{01}]$\ \&\ $[2,1_{10}]$ \\
    HH--J1.33--34 & 28.9  & $\aep$[0.0]       & $[0,1_{11}]$            & $[0,2_{11}]$            \\
    HH--J1.35--37 & 30.4  & $\fkp$[30.1]      & $[1,1_{11}]$            & $[2,0_{00}]$            \\
    HH--J1.38--43 & 30.6  & $\fkp$[4.2]       & $[1,1_{11}]$            & $[1,0_{00}]$\ \&\ $[2,0_{00}]$ \\
    HH--J1.44--46 & 31.9  & $\fkm$[31.6]      & $[1,1_{10}]$            & $[1,1_{01}]$\ \&\ $[2,1_{10}]$ \\
    HH--J1.47--49 & 32.0  & $\fem$[31.7]      & $[1,1_{10}]$            & $[1,1_{01}]$\ \&\ $[2,1_{10}]$ \\
    HH--J1.50--51 & 35.4  & $\em$[35.1]       & $[2,1_{10}]$            & ~                    \\
    HH--J1.52--54 & 35.8  & $\fep$[35.5]      & $[1,1_{11}]$            & ~                    \\
    HH--J1.55--56 & 36.0  & $\ep$[35.7]       & $[2,0_{00}]$            & $[2,1_{11}]$               \\
\hline\hline
\end{tabular}
\end{table}

\begin{table}[ht!]
\caption{%
  Rovibrational states of \metwatb\ (HD) with $J=1$. 
  See also caption to Table~\ref{tab:hhj1}.
  \label{tab:hdj1}
}
\begin{tabular}{@{}l@{\ \ }c@{\ \ }l@{\ \ }l@{\ \ }l@{}}
\hline\hline
\multicolumn{1}{l}{Label} & 
\multicolumn{1}{c}{$E$ [\cm]} & 
\multicolumn{1}{l}{Vib. parent} & 
\multicolumn{2}{c}{Coupled-rotor states ($J=1$)} \\ 
\cline{4-5}\\[-0.4cm]
 & & & \multicolumn{1}{l}{dominant} & \multicolumn{1}{l}{minor} \\
\hline
    HD--J1.1      & 0.3   & $\aep$[0.0]       & $[0,0_{00}]$ & $[0,2_{02}]$            \\
    HD--J1.2      & 1.3   & $\akm$[1.0]       & $[0,1_{01}]$ & $[3,1_{01}]$            \\
    HD--J1.3--5   & 3.3   & $\fkp$[3.0]       & $[1,0_{00}]$ & $[2,0_{00}]$\ \&\ $[1,2_{02}]$ \\
    HD--J1.6--8   & 4.2   & $\fem$[4.0]       & $[1,1_{01}]$ & ~                    \\
    HD--J1.9--14  & 6.9   & $\fkp$[20.4], $\fep$[20.6] & $[1,0_{00}]$ & $[1,2_{02}]$            \\
    HD--J1.15--20 & 7.9   & $\fem$[19.7], $\fkm$[19.7] & $[1,1_{01}]$ & ~                    \\
    HD--J1.21--22 & 10.1  & $\ep$[23.5]       & $[2,0_{00}]$ & ~                    \\
    HD--J1.23--24 & 11.0  & $\em$[22.7]       & $[2,1_{01}]$ & ~                    \\
    HD--J1.25--26 & 13.5  & $\akm$[1.0]       & $[0,1_{10}]$ & ~                    \\
    HD--J1.27--28 & 14.3  & $\aep$[0.0]       & $[0,1_{11}]$ & $[0,2_{11}]$            \\
    HD--J1.29--34 & 16.2  & $\fem$[4.0]       & $[1,1_{10}]$ & $[2,1_{10}]$            \\
    HD--J1.35--40 & 17.0  & $\fkp$[3.0]       & $[1,1_{11}]$ & $[2,1_{11}]$\ \&\ $[1,2_{11}]$ \\
    HD--J1.41--43 & 20.0  & $\fem$[19.7]      & $[1,1_{10}]$ & $[2,1_{10}]$            \\
    HD--J1.44--46 & 20.0  & $\fkm$[19.7]      & $[1,1_{10}]$ & $[2,1_{10}]$            \\
    HD--J1.47--49 & 20.7  & $\fkp$[20.4]      & $[1,1_{11}]$ & $[1,2_{11}]$            \\
    HD--J1.50--52 & 20.9  & $\fep$[20.6]      & $[1,1_{11}]$ & $[1,2_{11}]$            \\
    HD--J1.53--54 & 23.0  & $\em$[22.7]       & $[2,1_{10}]$ & ~                    \\
    HD--J1.55--56 & 23.8  & $\ep$[23.5]       & $[2,1_{11}]$ & ~                    \\
\hline\hline
\end{tabular}
\end{table}

\begin{table}[ht]
\caption{%
  Rovibrational states of \metwatc\ (DH) with $J=1$. 
  See also caption to Table~\ref{tab:hhj1}.
  \label{tab:dhj1}
}
\begin{tabular}{@{}l@{\ \ }c@{\ \ }l@{\ \ }l@{\ \ }l@{}}
\hline\hline
\multicolumn{1}{l}{Label} & 
\multicolumn{1}{c}{$E$ [\cm]} & 
\multicolumn{1}{l}{Vib. parent} & 
\multicolumn{2}{c}{Coupled-rotor states ($J=1$)} \\ 
\cline{4-5}\\[-0.4cm]
 & & & \multicolumn{1}{l}{dominant} & \multicolumn{1}{l}{minor} \\
\hline
    DH--J1.1      & 0.3   & $\aep$[0.0]                 & $[0,0_{00}]$            & ~                    \\

    DH--J1.2--4   & 1.0   & $\fkp$[0.8]                 & $[1,0_{00}]$            & $[2,0_{00}]$            \\
    DH--J1.5--6   & 3.2   & $\fkp$[27.2], $\fep$[28.9]  & $[1,0_{00}]$            & $[2,0_{00}]$            \\
    DH--J1.7--10  & 3.2   & $\ep$[26.4]\ \&\ & $[1,0_{00}]$            & ~       \\
    & & $\fkp$[27.2], $\fep$[28.9] & & \\
    DH--J1.11--12 & 3.9   & $\fkp$[27.2]                & $[2,0_{00}]$            & ~                    \\
    DH--J1.13     & 6.0   & $\akm$[5.7]                 & $[0,1_{01}]$            & $[3,1_{01}]$            \\
    DH--J1.14--16 & 6.7   & $\fem$[6.4]                 & $[1,1_{01}]$            & $[2,1_{01}]$            \\
    DH--J1.17--22 & 8.9   & $\fem$[26.0], $\fkm$[26.2]  & $[1,1_{01}]$            & $[2,1_{01}]$\ \&\ $[3,1_{01}]$ \\
    DH--J1.23--24 & 9.6   & $\em$[26.6]                 & $[2,1_{01}]$            & ~                    \\
    DH--J1.25--26 & 23.7  & $\akm$[5.7]                 & $[0,1_{10}]$            & ~                    \\
    DH--J1.27--32 & 24.1  & $\fem$[6.4]                 & $[1,1_{10}]$            & $[2,1_{10}]$            \\
    DH--J1.33--36 & 26.1  & $\fkp$[0.8]                 & $[0,1_{10}]$            & ~ \\
    DH--J1.37--38,41 & 26.3  & $\fem$[26.0]             & $[1,1_{10}]$            & $[3,1_{10}]$ \& $[2,1_{10}]$\\
    DH--J1.39--40 & 26.3  & $\fkp$[0.8]                 & $[0,1_{10}]$            & ~ \\
    DH--J1.41--44 & 26.4  & $\fkm$[26.2]                & $[1,1_{10}]$            & $[2,1_{10}]$ \\
    DH--J1.45--46 & 26.7  & $\aep$[0.0]                 & $[0,1_{11}]$            & ~                    \\
    DH--J1.47--48 & 26.7  & $\ep$[26.4]                 & $[1,0_{00}]$\ \&\ $[1,1_{11}]$ & $[2,0_{00}]$\ \&\ $[2,1_{11}]$ \\
    DH--J1.49--50 & 26.8  & $\em$[26.6]                 & $[2,1_{10}]$            & ~                    \\
    DH--J1.51--53 & 27.9  & $\fkp$[27.2]                & $[2,0_{00}]$\ \&\ $[1,1_{11}]$ & $[2,1_{11}]$            \\
    DH--J1.54--56 & 29.3  & $\fep$[28.9]                & $[1,1_{11}]$            & ~                    \\
\hline\hline
\end{tabular}
\end{table}

\begin{table}[ht]
\caption{%
  Rovibrational states of \metwatd\ (DD) with $J=1$. 
  See also caption to Table~\ref{tab:hhj1}.
  \label{tab:ddj1}
}
\begin{tabular}{@{}l@{\ \ }c@{\ \ }l@{\ \ }l@{\ \ }l@{}}
\hline\hline
\multicolumn{1}{l}{Label} & 
\multicolumn{1}{c}{$E$ [\cm]} & 
\multicolumn{1}{l}{Vib. parent} & 
\multicolumn{2}{c}{Coupled-rotor states ($J=1$)} \\ 
\cline{4-5}\\[-0.4cm]
 & & & \multicolumn{1}{l}{dominant} & \multicolumn{1}{l}{minor} \\
\hline
    DD--J1.1      & 0.2   & $\aep$[0.0]       & $[0,0_{00}]$ & $[0,2_{02}]$            \\
    DD--J1.2--4   & 0.8   & $\fkp$[0.5]       & $[1,0_{00}]$ & $[2,0_{00}]$\ \&\ $[1,2_{02}]$ \\
    DD--J1.5      & 1.1   & $\akm$[0.9]       & $[0,1_{01}]$ & $[2,1_{01}]$            \\
    DD--J1.6--8   & 1.6   & $\fem$[1.3]       & $[1,1_{01}]$ & $[2,1_{01}]$            \\
    DD--J1.9--14  & 3.0   & $\fkp$[16.2], $\fep$[16.3] & $[1,0_{00}]$ & $[2,0_{00}]$            \\
    DD--J1.15--16 & 3.5   & $\ep$[16.6]       & $[2,0_{00}]$ & ~                    \\
    DD--J1.17--22 & 3.9   & $\fem$[15.5], $\fkm$[15.5] & $[1,1_{01}]$ & $[2,1_{01}]$            \\
    DD--J1.23--24 & 4.4   & $\em$[15.9]       & $[2,1_{01}]$ & ~                    \\
    DD--J1.25--26 & 12.9  & $\akm$[0.9]       & $[0,1_{10}]$ & ~                    \\
    DD--J1.27--32 & 13.3  & $\fem$[1.3]       & $[1,1_{10}]$ & $[2,1_{10}]$            \\
    DD--J1.33--34 & 13.7  & $\aep$[0.0]       & $[0,1_{11}]$ & $[0,2_{11}]$            \\
    DD--J1.35--40 & 14.1  & $\fkp$[0.5]       & $[1,1_{11}]$ & $[2,1_{11}]$            \\
    DD--J1.41--43 & 15.7  & $\fem$[15.5]      & $[1,1_{10}]$ & $[2,1_{10}]$            \\
    DD--J1.44--46 & 15.7  & $\fkm$[15.5]      & $[1,1_{10}]$ & $[2,1_{10}]$            \\
    DD--J1.47--48 & 16.2  & $\em$[15.9]       & $[2,1_{10}]$ & ~                    \\
    DD--J1.49--51 & 16.4  & $\fkp$[16.2]      & $[1,1_{11}]$ & ~                    \\
    DD--J1.42--54 & 16.5  & $\fep$[16.3]      & $[1,1_{11}]$ & ~                    \\
    DD--J1.55--56 & 16.9  & $\ep$[16.6]       & $[2,1_{11}]$ & ~                    \\
\hline\hline
\end{tabular}
\end{table}


\begin{table}[ht!]
\caption{%
  Rovibrational states of \metwata\ (HH) with $J=2$. 
  See also caption to Table~\ref{tab:hhj1}.
  \label{tab:hhj2}
}
\begin{tabular}{@{}l@{\ \ }c@{\ \ }l@{\ \ }l@{\ \ }l@{}}
\hline\hline
\multicolumn{1}{l}{Label} & 
\multicolumn{1}{c}{$E$ [\cm]} & 
\multicolumn{1}{l}{Vib. parent} & 
\multicolumn{2}{c}{Coupled-rotor states ($J=2$)} \\ 
\cline{4-5}\\[-0.4cm]
 & & & \multicolumn{1}{l}{dominant} & \multicolumn{1}{l}{minor} \\
\hline
    HH--J2.1      & 0.9   & $\aep$[0.0]       & $[0,0_{00}]$            & ~                    \\
    HH--J2.2--4   & 5.0   & $\fkp$[4.2]       & $[1,0_{00}]$            & $[2,0_{00}]$            \\
    HH--J2.5      & 7.5   & $\akm$[6.6]       & $[0,1_{01}]$            & ~                    \\
    HH--J2.6--11  & 8.1   & $\fkp$[30.1], $\fep$[35.5] & $[1,0_{00}]$            & $[2,0_{00}]$            \\
    HH--J2.12--14 & 11.2  & $\fem$[10.3]      & $[1,1_{01}]$            & $[2,1_{01}]$            \\
    HH--J2.15--16 & 12.6  & $\ep$[35.7]       & $[2,0_{00}]$            & ~                    \\
    HH--J2.17--22 & 14.5  & $\fkm$[31.6], $\fem$[31.7] & $[1,1_{01}]$            & $[2,1_{01}]$\ \&\ $[3,1_{01}]$ \\
    HH--J2.23--24 & 18.5  & $\em$[35.1]       & $[2,1_{01}]$            & ~                    \\
    HH--J2.25--26 & 22.6  & many $\Gamma^+$ states & $[2,0_{00}]$            & ~                    \\
    HH--J2.27--32 & 25.4  & many $\Gamma^+$ states & $[2,0_{00}]$            & ~                    \\
    HH--J2.33--34 & 26.1  & $\akm$[6.6]       & $[0,1_{10}]$            & $[0,1_{01}]$            \\
    HH--J2.35--36 & 28.8  & many $\Gamma^-$ states & $[2,1_{01}]$            & ~                    \\
    HH--J2.37--42 & 29.2  & $\fem$[10.3]      & $[1,1_{10}]$            & $[1,1_{01}]$\ \&\ $[2,1_{10}]$ \\
    HH--J2.43--44 & 29.5  & $\aep$[0.0]       & $[0,1_{11}]$            & ~                    \\
    HH--J2.45--47 & 30.9  & $\fkp$[30.1]      & $[1,1_{11}]$            & $[1,0_{00}]$\ \&\ $[2,0_{00}]$ \\
    HH--J2.48--53 & 31.2  & $\fkp$[4.2]       & $[1,1_{11}]$            & $[2,0_{00}]$            \\
    HH--J2.54--59 & 31.6  & many $\Gamma^-$ states & $[2,1_{01}]$            & $[3,1_{01}]$            \\
    HH--J2.60--62 & 32.5  & $\fkm$[31.6]      & $[1,1_{10}]$            & $[1,1_{01}]$            \\
    HH--J2.63--65 & 32.6  & $\fem$[31.7]      & $[1,1_{10}]$            & $[1,1_{01}]$\ \&\ $[2,1_{10}]$ \\
    HH--J2.66--71 & 33.0  & $\fkm$[31.6], $\fem$[31.7] & $[1,1_{10}]$            & $[2,1_{10}]$            \\
    HH--J2.72--73 & 36.0  & $\em$[35.1]       & $[2,1_{10}]$            & ~                    \\
    HH--J2.74--75 & 36.1  & $\em$[35.1]       & $[2,1_{10}]$            & ~                    \\
    HH--J2.76--78 & 36.4  & $\fep$[35.5]      & $[1,1_{11}]$            & ~                    \\
    HH--J2.79--80 & 36.5  & $\ep$[35.7]       & $[2,0_{00}]$            & ~                    \\
    HH--J2.81--86 & 36.6  & $\fep$[35.5]      & $[1,1_{11}]$            & ~                    \\
    HH--J2.87     & 36.9  & $\aep$(SM)[36.1]  & $[0,0_{00}]$\ \&\ $[0,1_{11}]$ & $[3,0_{00}]$            \\
    HH--J2.88--89 & 38.2  & $\ep$[35.7]       & $[2,1_{11}]$            & ~                    \\
\hline\hline
\end{tabular}
\end{table}

\begin{table}[ht]
\caption{%
  Rovibrational states of \metwatb\ (HD) with $J=2$. 
  See also caption to Table~\ref{tab:hhj1}.
  \label{tab:hdj2}
}
\begin{tabular}{@{}l@{\ \ }c@{\ \ }l@{\ \ }l@{\ \ }l@{}}
\hline\hline
\multicolumn{1}{l}{Label} & 
\multicolumn{1}{c}{$E$ [\cm]} & 
\multicolumn{1}{l}{Vib. parent} & 
\multicolumn{2}{c}{Coupled-rotor states ($J=2$)} \\ 
\cline{4-5}\\[-0.4cm]
 & & & \multicolumn{1}{l}{dominant} & \multicolumn{1}{l}{minor} \\
\hline
    HD--J2.1      & 0.8  & $\aep$[0.0]       & $[0,0_{00}]$ & $[0,2_{02}]$              \\
    HD--J2.2      & 1.9  & $\akm$[1.0]       & $[0,1_{01}]$ & ~                      \\
    HD--J2.3--5   & 3.8  & $\fkp$[3.0]       & $[1,0_{00}]$ & $[2,0_{00}]$\ \&\ $[1,2_{02}]$   \\
    HD--J2.6--8   & 4.8  & $\fem$[4.0]       & $[1,1_{01}]$ & $[2,1_{01}]$              \\
    HD--J2.9--14  & 7.4  & $\fkp$[20.4], $\fep$[20.6] & $[1,0_{00}]$ & $[1,2_{02}]$              \\
    HD--J2.15--20 & 8.4  & $\fem$[19.7], $\fkm$[19.7] & $[1,1_{01}]$ & $[2,1_{01}]$              \\
    HD--J2.21--22 & 10.7 & $\ep$[22.7]       & $[2,0_{00}]$ & ~                      \\
    HD--J2.23--24 & 11.6 & $\em$[23.5]       & $[2,1_{01}]$ & ~                      \\
    HD--J2.25--26 & 14.0 & $\akm$[1.0]       & $[0,1_{10}]$ & ~                      \\
    HD--J2.27--28 & 14.9 & $\aep$[0.0]       & $[0,11_{1}]$ & $[0,2_{11}]$              \\
    HD--J2.29--34 & 16.8 & $\fem$[4.0]       & $[1,1_{10}]$ & $[2,1_{10}]$              \\
    HD--J2.35--40 & 17.6 & $\fkp$[3.0]       & $[1,1_{11}]$ & $[2,1_{11}]$ \& $[1,2_{11}]$ \\
    HD--J2.41--46 & 20.4 & $\fem$[19.7], $\fkm$[19.7] & $[1,1_{10}]$ & ~                      \\
    HD--J2.47--49 & 20.6 & $\fem$[19.7]      & $[1,1_{10}]$ & $[2,1_{10}]$              \\
    HD--J2.50--52 & 20.6 & $\fkm$[19.7]      & $[1,1_{10}]$ & $[2,1_{10}]$              \\
    HD--J2.53--54,57 & 21.2 & $\fkp$[20.4]   & $[1,1_{11}]$ & \\
    HD--J2.55--56,58 & 21.2 & $\fkp$[20.4], $\fep$[20.6] & $[1,1_{11}]$ & $[1,2_{11}]$              \\

    HD--J2.59--61 & 21.3 & $\fkp$[20.4], $\fep$[20.6] & $[1,1_{11}]$ & $[1,2_{11}]$              \\
    HD--J2.62--64 & 21.5 & $\fep$[20.6]      & $[1,1_{11}]$ & $[1,2_{11}]$              \\

    HD--J2.65--66 & 21.8 & many $\Gamma^+$ states & $[2,0_{00}]$ & ~                      \\
    HD--J2.67--68 & 22.8 & many $\Gamma^-$ states & $[2,1_{01}]$ & ~                      \\
    HD--J2.69--72 & 23.6 & $\em$[23.5]       & $[2,1_{10}]$ & ~                      \\
    HD--J2.73--74 & 24.1 & $\ep$[22.7]       & $[2,1_{11}]$ & ~                      \\
    HD--J2.75--76 & 24.3 & $\ep$[22.7]       & $[2,1_{11}]$ & ~                      \\
    HD--J2.77--82 & 24.3 & many $\Gamma^+$ states & $[2,0_{00}]$ & $[2,2_{02}]$              \\
    HD--J2.83--88 & 25.3 & many $\Gamma^-$ states & $[2,1_{01}]$ & ~                      \\
\hline\hline
\end{tabular}
\end{table}

\begin{table}[ht]
\caption{%
  Rovibrational states of \metwatc\ (DH) with $J=2$. 
  See also caption to Table~\ref{tab:hhj1}.
  \label{tab:dhj2}
}
\begin{tabular}{@{}l@{\ \ }c@{\ \ }l@{\ \ }l@{\ \ }l@{}}
\hline\hline
\multicolumn{1}{l}{Label} & 
\multicolumn{1}{c}{$E$ [\cm]} & 
\multicolumn{1}{l}{Vib. parent} & 
\multicolumn{2}{c}{Coupled-rotor states ($J=2$)} \\ 
\cline{4-5}\\[-0.4cm]
 & & & \multicolumn{1}{l}{dominant} & \multicolumn{1}{l}{minor} \\
\hline
    DH--J2.1      & 0.8   & $\aep$[0.0]                 & $[0,0_{00}]$            & ~                    \\
    DH--J2.2--4   & 1.5   & $\fkp$[0.8]                 & $[1,0_{00}]$            & $[2,0_{00}]$            \\
    DH--J2.5--6   & 3.7   & $\fkp$[27.2], $\fep$[28.9]  & $[1,0_{00}]$            & $[2,0_{00}]$            \\
    DH--J2.7--10  & 3.7   & $\ep$[26.4]\ \&\ & $[1,0_{00}]$            & ~       \\
     & & $\fkp$[27.2], $\fep$[28.9] & & \\
    DH--J2.11--12 & 4.4   & $\fkp$[27.2]   & $[2,0_{00}]$            & ~                    \\
    DH--J2.13     & 6.5   & $\akm$[5.7]                 & $[0,1_{01}]$            & $[3,1_{01}]$            \\
    DH--J2.14--16 & 7.2   & $\fem$[6.4]                 & $[1,1_{01}]$            & $[2,1_{01}]$            \\
    DH--J2.17--22 & 9.4   & $\fem$[26.0], $\fkm$[26.2]  & $[1,1_{01}]$            & $[2,1_{01}]$, $[3,1_{01}]$ \\
    DH--J2.23--24 & 10.1  & $\em$[26.6]                 & $[2,1_{01}]$            & ~                    \\
    DH--J2.25--26 & 11.0  & many $\Gamma^+$ states  & $[2,0_{00}]$            & ~                    \\
    DH--J2.27--32 & 11.6  & many $\Gamma^+$ states  & $[2,0_{00}]$            & $[4,0_{00}]$            \\
    DH--J2.33--34 & 16.7  & many $\Gamma^-$ states  & $[2,1_{01}]$            & ~                    \\
    DH--J2.35--40 & 17.3  & many $\Gamma^-$ states  & $[2,1_{01}]$            & $[3,1_{01}]$            \\
    DH--J2.41--42 & 24.2  & $\akm$[5.7]                 & $[0,1_{10}]$            & ~                    \\
    DH--J2.43--48 & 24.6  & $\fem$[6.4]                 & $[1,1_{10}]$\ \&\ $[2,1_{10}]$ & ~                    \\
    DH--J2.49--52 & 26.6  & $\fkp$[0.8]                 & $[1,1_{11}]$            & $[2,0_{00}]$            \\
    DH--J2.53--54,57 & 26.8  & $\fem$[26.0]             & $[1,1_{10}]$            & ~                    \\
    DH--J2.55--56 & 26.8  & $\fkp$[0.8]                 & $[1,1_{11}]$            & $[2,0_{00}]$            \\
    DH--J2.57--60,64 & 26.9  & $\fkm$[26.2]             & $[1,1_{10}]$            & $[2,1_{10}]$            \\
    DH--J2.62--63 & 27.0  & $\fem$[26.0], $\fkm$[26.2]  & $[1,1_{10}]$            & $[2,1_{10}]$            \\
    DH--J2.65--66 & 27.0  & $\fem$[26.0], $\fkm$[26.2]  & $[1,1_{10}]$            & $[2,1_{10}]$            \\
		DH--J2.67--68 & 27.2  & $\aep$[0.0]                 & $[0,1_{11}]$            & ~                    \\
    DH--J2.69--70 & 27.2  & $\ep$[26.4]                 & $[1,0_{00}]$            & $[2,0_{00}]$            \\
    DH--J2.71--72 & 27.3  & $\em$[26.6]                 & $[2,1_{10}]$            & ~                    \\
    DH--J2.73--74 & 27.3  & $\em$[26.6]                 & $[2,1_{10}]$            & ~                    \\
    DH--J2.75--77 & 28.4  & $\fkp$[27.2]   & $[2,0_{00}]$            & $[1,1_{11}]$\ \&\ $[2,1_{11}]$ \\
    DH--J2.78--79 & 29.4  & $\fkp$[27.2]                & $[2,1_{11}]$            & ~       \\
    DH--J2.80--82 & 29.8  & $\fep$[28.9]                & $[1,1_{11}]$            & $[2,1_{11}]$            \\
    DH--J2.83--88 & 30.1  & $\fkp$[27.2], $\fep$[28.9]  & $[1,1_{11}]$            & ~                    \\
\hline\hline
\end{tabular}
\end{table}

\begin{table}[ht]
\caption{%
  Rovibrational states of \metwatd\ (DD) with $J=2$. 
  See also caption to Table~\ref{tab:hhj1}.
  \label{tab:ddj2}
}
\begin{tabular}{@{}l@{\ \ }c@{\ \ }l@{\ \ }l@{\ \ }l@{}}
\hline\hline
\multicolumn{1}{l}{Label} & 
\multicolumn{1}{c}{$E$ [\cm]} & 
\multicolumn{1}{l}{Vib. parent} & 
\multicolumn{2}{c}{Coupled-rotor states ($J=2$)} \\ 
\cline{4-5}\\[-0.4cm]
 & & & \multicolumn{1}{l}{dominant} & \multicolumn{1}{l}{minor} \\
\hline
    DD--J2.1     & 0.7    & $\aep$[0.0]        & $[0,0_{00}]$            & $[0,2_{02}]$            \\
    DD--J2.2--4  & 1.2    & $\fkp$[0.5]        & $[1,0_{00}]$            & $[2,0_{00}]$\ \&\ $[1,2_{02}]$ \\
    DD--J2.5     & 1.6    & $\akm$[0.9]        & $[0,1_{01}]$            & $[3,1_{01}]$            \\
    DD--J2.6--8  & 2.1    & $\fem$[1.3]        & $[1,1_{01}]$            & $[2,1_{01}]$            \\
    DD--J2.9--14 & 3.5    & $\fkp$[16.2], $\fep$[16.3] & $[1,0_{00}]$            & $[2,0_{00}]$            \\
    DD--J2.15--16 & 4.0   & $\ep$[16.6]        & $[2,0_{00}]$            & ~                    \\
    DD--J2.17--22 & 4.4   & $\fem$[15.5], $\fkm$[15.5] & $[1,1_{01}]$            & $[1,1_{10}]$\ \&\ $[3,1_{01}]$ \\
    DD--J2.23--24 & 4.8   & $\em$[15.9]        & $[2,1_{01}]$            & ~                    \\
    DD--J2.25--26 & 10.9  & many $\Gamma^+$ states  & $[2,0_{00}]$            & ~                    \\
    DD--J2.27--32 & 11.3  & many $\Gamma^+$ states  & $[2,0_{00}]$            & $[3,0_{00}]$            \\
    DD--J2.33--34 & 11.7  & many $\Gamma^-$ states  & $[2,1_{01}]$            & ~                    \\
    DD--J2.35--40 & 12.2  & many $\Gamma^-$ states  & $[2,1_{01}]$\ \&\ $[3,1_{01}]$ & ~                    \\
    DD--J2.41--42 & 13.4  & $\akm$[0.9]        & $[0,1_{10}]$            & ~                    \\
    DD--J2.43--48 & 13.8  & $\fem$[1.3]        & $[1,1_{10}]$            & $[2,1_{10}]$            \\
    DD--J2.49--50 & 14.2  & $\aep$[0.0]        & $[0,1_{11}]$            & $[0,2_{11}]$            \\
    DD--J2.51--56 & 14.6  & $\fkp$[0.5]        & $[1,1_{11}]$            & ~                    \\
    DD--J2.57--62 & 16.0  & $\fem$[15.5], $\fkm$[15.5] & $[1,1_{10}]$            & $[2,1_{10}]$            \\
    DD--J2.63--65 & 16.3  & $\fem$[15.5]       & $[1,1_{10}]$            & $[2,1_{10}]$            \\
    DD--J2.66--68 & 16.3  & $\fkm$[15.5]       & $[1,1_{10}]$            & $[2,1_{10}]$            \\
    DD--J2.69--70 & 16.4  & $\em$[15.9]        & $[2,1_{10}]$            & ~                    \\
    DD--J2.71--72 & 16.6  & $\em$[15.9]        & $[2,1_{10}]$            & ~                    \\
    DD--J2.73--78 & 16.8  & $\fkp$[16.2], $\fep$[16.3] & $[1,1_{11}]$            & ~                    \\
    DD--J2.79--81 & 17.0  & $\fkp$[16.2]       & $[1,1_{11}]$            & ~                    \\
    DD--J2.82--84 & 17.0  & $\fep$[16.3]       & $[1,1_{11}]$            & ~                    \\
    DD--J2.85--86 & 17.2  & $\ep$[16.6]        & $[2,1_{11}]$            & ~                    \\
    DD--J2.87--88 & 17.4  & $\ep$[16.6]        & $[2,1_{11}]$            & ~                    \\
\hline\hline
\end{tabular}
\end{table}

\newpage

\begin{table}[ht!]
\caption{%
  Sum of the energies of the methane and the water rotors in \metwata\ (HH).
  \label{tab:sumrotHH}
}
\begin{tabular}{c@{\ \ }c@{\ \ }c@{\ \ }c|@{\ \ }r@{\ \ }r@{\ \ }r@{\ \ }r@{\ \ }r@{\ \ }r@{\ \ }r@{}}
\hline\hline
    \multicolumn{3}{l}{\metwata}   & $\jmet$ & 0     & 1     & 2     & 3     & 4     & 5     & 6   \\ [0.1cm]
    $\jwat$ & $\ka$ & $\kc$ &E\ [\cm] & 0.0   & 10.2  & 30.5  & 61.1  & 101.8 & 152.7 & 213.8 \\ [0.1cm] 
		\hline
    0          & 0  & 0 & 0.0      & 0.0   & 10.2  & 30.5  & 61.1  & 101.8 & 152.7 & 213.8 \\
    1          & 0  & 1 & 23.2     & 23.2  & 33.4  & 53.8  & 84.3  & 125.1 & 176.0 & 237.0 \\
    1          & 1  & 1 & 35.5     & 35.5  & 45.7  & 66.1  & 96.6  & 137.3 & 188.2 & 249.3 \\
    1          & 1  & 0 & 40.4     & 40.4  & 50.6  & 71.0  & 101.5 & 142.2 & 193.1 & 254.2 \\
    2          & 0  & 2 & 68.5     & 68.5  & 78.7  & 99.1  & 129.6 & 170.3 & 221.2 & 282.3 \\
    2          & 1  & 2 & 77.1     & 77.1  & 87.3  & 107.7 & 138.2 & 178.9 & 229.8 & 290.9 \\
    2          & 1  & 1 & 91.8     & 91.8  & 102.0 & 122.3 & 152.9 & 193.6 & 244.5 & 305.6 \\
    2          & 2  & 1 & 128.7    & 128.7 & 138.8 & 159.2 & 189.7 & 230.5 & 281.4 & 342.5 \\
    2          & 2  & 0 & 129.8    & 129.8 & 140.0 & 160.4 & 190.9 & 231.7 & 282.6 & 343.7 \\
    3          & 0  & 3 & 133.9    & 133.9 & 144.0 & 164.4 & 194.9 & 235.7 & 286.6 & 347.7 \\
    3          & 1  & 3 & 138.8    & 138.8 & 149.0 & 169.3 & 199.9 & 240.6 & 291.5 & 352.6 \\
    3          & 1  & 2 & 168.0    & 168.0 & 178.2 & 198.5 & 229.1 & 269.8 & 320.7 & 381.8 \\
    3          & 2  & 2 & 198.4    & 198.4 & 208.5 & 228.9 & 259.5 & 300.2 & 351.1 & 412.2 \\
    3          & 2  & 1 & 203.9    & 203.9 & 214.1 & 234.5 & 265.0 & 305.8 & 356.7 & 417.8 \\
    3          & 3  & 1 & 272.7    & 272.7 & 282.9 & 303.3 & 333.8 & 374.5 & 425.4 & 486.5 \\
    3          & 3  & 0 & 272.9    & 272.9 & 283.1 & 303.4 & 334.0 & 374.7 & 425.6 & 486.7 \\
		\hline\hline
\end{tabular}
\end{table}

\begin{table}
\caption{%
  Sum of the energies of the methane and the water rotors in \metwatb\ (HD).
  \label{tab:sumrotHD}
}
\begin{tabular}{c@{\ \ }c@{\ \ }c@{\ \ }c|@{\ \ }r@{\ \ }r@{\ \ }r@{\ \ }r@{\ \ }r@{\ \ }r@{\ \ }r@{}}
\hline\hline
    \multicolumn{3}{l}{\metwatb}  & $\jmet$ & 0     & 1     & 2     & 3     & 4     & 5     & 6   \\ [0.1cm]
		$\jwat$ & $\ka$ & $\kc$ &E\ [\cm]& 0.0   & 10.2  & 30.5  & 61.1  & 101.8 & 152.7 & 213.8 \\ [0.1cm] 
		\hline
		0          & 0  & 0 & 0.0     & 0.0   & 10.2  & 30.5  & 61.1  & 101.8 & 152.7 & 213.8 \\
    1          & 0  & 1 & 11.9    & 11.9  & 22.1  & 42.5  & 73.0  & 113.7 & 164.6 & 225.7 \\
    1          & 1  & 1 & 19.6    & 19.6  & 29.8  & 50.1  & 80.7  & 121.4 & 172.3 & 233.4 \\
    1          & 1  & 0 & 21.9    & 21.9  & 32.1  & 52.5  & 83.0  & 123.7 & 174.6 & 235.7 \\
    2          & 0  & 2 & 35.3    & 35.3  & 45.5  & 65.8  & 96.4  & 137.1 & 188.0 & 249.1 \\
    2          & 1  & 2 & 41.1    & 41.1  & 51.3  & 71.7  & 102.2 & 142.9 & 193.8 & 254.9 \\
    2          & 1  & 1 & 48.1    & 48.1  & 58.2  & 78.6  & 109.1 & 149.9 & 200.8 & 261.9 \\
    2          & 2  & 1 & 71.1    & 71.1  & 81.3  & 101.7 & 132.2 & 172.9 & 223.8 & 284.9 \\
    2          & 2  & 0 & 71.6    & 71.6  & 81.7  & 102.1 & 132.7 & 173.4 & 224.3 & 285.4 \\
    3          & 0  & 3 & 69.4    & 69.4  & 79.5  & 99.9  & 130.5 & 171.2 & 222.1 & 283.2 \\
    3          & 1  & 3 & 73.2    & 73.2  & 83.3  & 103.7 & 134.2 & 175.0 & 225.9 & 287.0 \\
    3          & 1  & 2 & 86.9    & 86.9  & 97.1  & 117.5 & 148.0 & 188.8 & 239.7 & 300.8 \\
    3          & 2  & 2 & 106.9   & 106.9 & 117.0 & 137.4 & 168.0 & 208.7 & 259.6 & 320.7 \\
    3          & 2  & 1 & 109.0   & 109.0 & 119.2 & 139.5 & 170.1 & 210.8 & 261.7 & 322.8 \\
    3          & 3  & 1 & 151.3   & 151.3 & 161.5 & 181.9 & 212.4 & 253.2 & 304.1 & 365.1 \\
    3          & 3  & 0 & 151.4   & 151.4 & 161.6 & 181.9 & 212.5 & 253.2 & 304.1 & 365.2 \\
		\hline\hline
\end{tabular}
\end{table}

\begin{table}[ht!]
\caption{%
  Sum of the energies of the methane and the water rotors in \metwatc\ (DH).
  \label{tab:sumrotDH}
}
\begin{tabular}{c@{\ \ }c@{\ \ }c@{\ \ }c|@{\ \ }r@{\ \ }r@{\ \ }r@{\ \ }r@{\ \ }r@{\ \ }r@{\ \ }r@{}}
\hline\hline
    \multicolumn{3}{l}{\metwatc}    & $\jmet$ & 0     & 1     & 2     & 3     & 4     & 5     & 6   \\ [0.1cm]
    $\jwat$ & $\ka$ & $\kc$ & E\ [\cm] & 0.0   & 5.1   & 15.4  & 30.9  & 51.5  & 77.2  & 108.1 \\ [0.1cm] 
		\hline
    0          & 0  & 0  & 0.0      & 0.0   & 5.1   & 15.4  & 30.9  & 51.5  & 77.2  & 108.1 \\
    1          & 0  & 1  & 23.2     & 23.2  & 28.4  & 38.7  & 54.1  & 74.7  & 100.4 & 131.3 \\
    1          & 1  & 1  & 35.5     & 35.5  & 40.7  & 51.0  & 66.4  & 87.0  & 112.7 & 143.6 \\
    1          & 1  & 0  & 40.4     & 40.4  & 45.6  & 55.9  & 71.3  & 91.9  & 117.6 & 148.5 \\
    2          & 0  & 2  & 68.5     & 68.5  & 73.7  & 84.0  & 99.4  & 120.0 & 145.7 & 176.6 \\
    2          & 1  & 2  & 77.1     & 77.1  & 82.3  & 92.5  & 108.0 & 128.6 & 154.3 & 185.2 \\
    2          & 1  & 1  & 91.8     & 91.8  & 96.9  & 107.2 & 122.7 & 143.3 & 169.0 & 199.9 \\
    2          & 2  & 1  & 128.7    & 128.7 & 133.8 & 144.1 & 159.5 & 180.1 & 205.8 & 236.7 \\
    2          & 2  & 0  & 129.8    & 129.8 & 135.0 & 145.3 & 160.7 & 181.3 & 207.0 & 237.9 \\
    3          & 0  & 3  & 133.9    & 133.9 & 139.0 & 149.3 & 164.7 & 185.3 & 211.0 & 241.9 \\
    3          & 1  & 3  & 138.8    & 138.8 & 143.9 & 154.2 & 169.7 & 190.3 & 216.0 & 246.9 \\
    3          & 1  & 2  & 168.0    & 168.0 & 173.1 & 183.4 & 198.9 & 219.4 & 245.2 & 276.1 \\
    3          & 2  & 2  & 198.4    & 198.4 & 203.5 & 213.8 & 229.2 & 249.8 & 275.6 & 306.4 \\
    3          & 2  & 1  & 203.9    & 203.9 & 209.1 & 219.4 & 234.8 & 255.4 & 281.1 & 312.0 \\
    3          & 3  & 1  & 272.7    & 272.7 & 277.9 & 288.2 & 303.6 & 324.2 & 349.9 & 380.8 \\
    3          & 3  & 0  & 272.9    & 272.9 & 278.1 & 288.3 & 303.8 & 324.4 & 350.1 & 381.0 \\
		\hline\hline
\end{tabular}
\end{table}

\begin{table}[ht!]
\caption{%
  Sum of the energies of the methane and the water rotors in \metwatd\ (DD).
  \label{tab:sumrotDD}
}
\begin{tabular}{c@{\ \ }c@{\ \ }c@{\ \ }c|@{\ \ }r@{\ \ }r@{\ \ }r@{\ \ }r@{\ \ }r@{\ \ }r@{\ \ }r@{}}
\hline\hline
    \multicolumn{3}{l}{\metwatd}   & $\jmet$ & 0     & 1     & 2     & 3     & 4     & 5     & 6   \\ [0.1cm]
    $\jwat$ & $\ka$ & $\kc$ &E\ [\cm] & 0.0   & 5.1   & 15.4  & 30.9  & 51.5  & 77.2  & 108.1 \\ [0.1cm] 
		\hline
    0          & 0  & 0  & 0.0     & 0.0   & 5.1   & 15.4  & 30.9  & 51.5  & 77.2  & 108.1 \\
    1          & 0  & 1  & 11.9    & 11.9  & 17.1  & 27.4  & 42.8  & 63.4  & 89.1  & 120.0 \\
    1          & 1  & 1  & 19.6    & 19.6  & 24.8  & 35.0  & 50.5  & 71.1  & 96.8  & 127.7 \\
    1          & 1  & 0  & 21.9    & 21.9  & 27.1  & 37.4  & 52.8  & 73.4  & 99.1  & 130.0 \\
    2          & 0  & 2  & 35.3    & 35.3  & 40.4  & 50.7  & 66.2  & 86.8  & 112.5 & 143.4 \\
    2          & 1  & 2  & 41.1    & 41.1  & 46.3  & 56.6  & 72.0  & 92.6  & 118.3 & 149.2 \\
    2          & 1  & 1  & 48.1    & 48.1  & 53.2  & 63.5  & 78.9  & 99.5  & 125.2 & 156.1 \\
    2          & 2  & 1  & 71.1    & 71.1  & 76.3  & 86.6  & 102.0 & 122.6 & 148.3 & 179.2 \\
    2          & 2  & 0  & 71.6    & 71.6  & 76.7  & 87.0  & 102.4 & 123.0 & 148.8 & 179.6 \\
    3          & 0  & 3  & 69.4    & 69.4  & 74.5  & 84.8  & 100.2 & 120.8 & 146.6 & 177.4 \\
    3          & 1  & 3  & 73.2    & 73.2  & 78.3  & 88.6  & 104.0 & 124.6 & 150.3 & 181.2 \\
    3          & 1  & 2  & 86.9    & 86.9  & 92.1  & 102.4 & 117.8 & 138.4 & 164.1 & 195.0 \\
    3          & 2  & 2  & 106.9   & 106.9 & 112.0 & 122.3 & 137.7 & 158.3 & 184.1 & 214.9 \\
    3          & 2  & 1  & 109.0   & 109.0 & 114.1 & 124.4 & 139.9 & 160.5 & 186.2 & 217.1 \\
    3          & 3  & 1  & 151.3   & 151.3 & 156.5 & 166.8 & 182.2 & 202.8 & 228.5 & 259.4 \\
    3          & 3  & 0  & 151.4   & 151.4 & 156.5 & 166.8 & 182.3 & 202.9 & 228.6 & 259.5 \\
\hline\hline
\end{tabular}
\end{table}

\begin{table} 
\caption{%
  Coupling energy, $\Eint$ defined in Section~III, of the rotors for 
  the studied methane-water isotopologues with $J=0$. 
  The second column contains the analytic solution in $2\mu R^2$ units, 
  the remaining columns give the numerical value of $\Eint$ in \cm\ for
  the equilibrium distance, $R=R_{\text{eq}}=6.5$~bohr, and for an arbitrarily chosen large value, $R=100$~bohr.
  The reduced masses are $\muhh=8.481697$~u, $\muhd=8.903115$~u, 
  $\mudh=9.489256$~u, and $\mudd=10.01987$~u.
  \label{tab:crf1}  
  }
  \begin{tabular}{@{}r@{\ \ }c@{\ \ }c@{\ \ }c@{\ \ }c@{\ \ }c@{\ \ }c@{\ \ }c@{\ \ }c@{\ \ }c@{}}
    \hline\hline
    \multicolumn{2}{c}{$J=0$} & 
    \multicolumn{2}{c}{\metwata} &
    \multicolumn{2}{c}{\metwatb} & 
    \multicolumn{2}{c}{\metwatc} & 
    \multicolumn{2}{c}{\metwatd}\\ 
    \cmidrule(lr){1-2} \cmidrule(lr){3-4}\cmidrule(lr){5-6}\cmidrule(lr){7-8}\cmidrule(lr){9-10}
    $j$ & 
    $\Eint\cdot 2\mu R^2$ & 
    $R_{\rm eq}$ & $R_{\rm 100}$ & 
    $R_{\rm eq}$ & $R_{\rm 100}$ &
    $R_{\rm eq}$ & $R_{\rm 100}$ &
    $R_{\rm eq}$ & $R_{\rm 100}$ \\
    \hline
    0 & 0  & 0.000       & 0.000       & 0.000       & 0.000       & 0.000       & 0.000       & 0.000       & 0.000       \\
    1 & 2  & 0.336       & 0.001       & 0.320       & 0.001       & 0.300       & 0.001       & 0.284       & 0.001       \\
    2 & 6  & 1.008       & 0.004       & 0.960       & 0.004       & 0.901       & 0.004       & 0.853       & 0.004       \\
    3 & 12 & 2.016       & 0.009       & 1.920       & 0.008       & 1.802       & 0.008       & 1.706       & 0.007       \\
    4 & 20 & 3.360       & 0.014       & 3.201       & 0.014       & 3.003       & 0.013       & 2.844       & 0.012       \\
    5 & 30 & 5.040       & 0.021       & 4.801       & 0.020       & 4.505       & 0.019       & 4.266       & 0.018       \\
  \hline\hline
\end{tabular}
\end{table}

\begin{table} 
\caption{%
  Coupling energy, $E_\text{c}$ in \cm, for $J=1$. See also caption to Table~\ref{tab:crf1}.
  \label{tab:crf2}
}
  \begin{tabular}{@{}r@{\ \ }c@{\ \ }c@{\ \ }c@{\ \ }c@{\ \ }c@{\ \ }c@{\ \ }c@{\ \ }c@{\ \ }c@{}}
    \hline\hline
    \multicolumn{2}{c}{$J=1$} &
    \multicolumn{2}{c}{\metwata} &
    \multicolumn{2}{c}{\metwatb} &
    \multicolumn{2}{c}{\metwatc} &
    \multicolumn{2}{c}{\metwatd}\\ 
    \cmidrule(lr){1-2} \cmidrule(lr){3-4}\cmidrule(lr){5-6}\cmidrule(lr){7-8}\cmidrule(lr){9-10}
    $j$ & 
    $\Eint\cdot 2\mu R^2$ & 
    $R_{\rm eq}$ & $R_{\rm 100}$ & 
    $R_{\rm eq}$ & $R_{\rm 100}$ &
    $R_{\rm eq}$ & $R_{\rm 100}$ &
    $R_{\rm eq}$ & $R_{\rm 100}$ \\
    \hline
    0 & 2  & 0.336       & 0.001       & 0.320       & 0.001       & 0.300       & 0.001       & 0.284       & 0.001       \\
    1 & 6  & 1.008       & 0.004       & 0.960       & 0.004       & 0.901       & 0.004       & 0.853       & 0.004       \\
    ~ & 2  & 0.336       & 0.001       & 0.320       & 0.001       & 0.300       & 0.001       & 0.284       & 0.001       \\
    ~ & 0  & 0.000       & 0.000       & 0.000       & 0.000       & 0.000       & 0.000       & 0.000       & 0.000       \\
    2 & 12 & 2.016       & 0.009       & 1.920       & 0.008       & 1.802       & 0.008       & 1.706       & 0.007       \\
    ~ & 6  & 1.008       & 0.004       & 0.960       & 0.004       & 0.901       & 0.004       & 0.853       & 0.004       \\
    ~ & 2  & 0.336       & 0.001       & 0.320       & 0.001       & 0.300       & 0.001       & 0.284       & 0.001       \\
    3 & 20 & 3.360       & 0.014       & 3.201       & 0.014       & 3.003       & 0.013       & 2.844       & 0.012       \\
    ~ & 12 & 2.016       & 0.009       & 1.920       & 0.008       & 1.802       & 0.008       & 1.706       & 0.007       \\
    ~ & 6  & 1.008       & 0.004       & 0.960       & 0.004       & 0.901       & 0.004       & 0.853       & 0.004       \\
    4 & 30 & 5.040       & 0.021       & 4.801       & 0.020       & 4.505       & 0.019       & 4.266       & 0.018       \\
    ~ & 20 & 3.360       & 0.014       & 3.201       & 0.014       & 3.003       & 0.013       & 2.844       & 0.012       \\
    ~ & 12 & 2.016       & 0.009       & 1.920       & 0.008       & 1.802       & 0.008       & 1.706       & 0.007       \\
    5 & 42 & 7.056       & 0.030       & 6.722       & 0.028       & 6.306       & 0.027       & 5.972       & 0.025       \\
    ~ & 30 & 5.040       & 0.021       & 4.801       & 0.020       & 4.505       & 0.019       & 4.266       & 0.018       \\
    ~ & 20 & 3.360       & 0.014       & 3.201       & 0.014       & 3.003       & 0.013       & 2.844       & 0.012       \\
\hline\hline
\end{tabular}
\end{table}

\begin{table}
\caption{%
  Coupling energy, $E_\text{c}$ in \cm, for $J=2$. See also caption to Table~\ref{tab:crf1}.
  \label{tab:crf3}
}
  \begin{tabular}{@{}r@{\ \ }c@{\ \ }c@{\ \ }c@{\ \ }c@{\ \ }c@{\ \ }c@{\ \ }c@{\ \ }c@{\ \ }c@{}}
    \hline\hline
    \multicolumn{2}{c}{$J=2$} &
    \multicolumn{2}{c}{\metwata} &
    \multicolumn{2}{c}{\metwatb} &
    \multicolumn{2}{c}{\metwatc} &
    \multicolumn{2}{c}{\metwatd}\\ 
    \cmidrule(lr){1-2} \cmidrule(lr){3-4}\cmidrule(lr){5-6}\cmidrule(lr){7-8}\cmidrule(lr){9-10}
    $j$ & 
    $\Eint\cdot 2\mu R^2$ & 
    $R_{\rm eq}$ & $R_{\rm 100}$ & 
    $R_{\rm eq}$ & $R_{\rm 100}$ &
    $R_{\rm eq}$ & $R_{\rm 100}$ &
    $R_{\rm eq}$ & $R_{\rm 100}$ \\
    \hline
    0 & 6  & 1.008       & 0.004       & 0.960       & 0.004       & 0.901       & 0.004       & 0.853       & 0.004       \\
    1 & 12 & 2.016       & 0.009       & 1.920       & 0.008       & 1.802       & 0.008       & 1.706       & 0.007       \\
    ~ & 6  & 1.008       & 0.004       & 0.960       & 0.004       & 0.901       & 0.004       & 0.853       & 0.004       \\
    ~ & 2  & 0.336       & 0.001       & 0.320       & 0.001       & 0.300       & 0.001       & 0.284       & 0.001       \\
    2 & 20 & 3.360       & 0.014       & 3.201       & 0.014       & 3.003       & 0.013       & 2.844       & 0.012       \\
    ~ & 12 & 2.016       & 0.009       & 1.920       & 0.008       & 1.802       & 0.008       & 1.706       & 0.007       \\
    ~ & 6  & 1.008       & 0.004       & 0.960       & 0.004       & 0.901       & 0.004       & 0.853       & 0.004       \\
    ~ & 2  & 0.336       & 0.001       & 0.320       & 0.001       & 0.300       & 0.001       & 0.284       & 0.001       \\
    ~ & 0  & 0.000       & 0.000       & 0.000       & 0.000       & 0.000       & 0.000       & 0.000       & 0.000       \\
    3 & 30 & 5.040       & 0.021       & 4.801       & 0.020       & 4.505       & 0.019       & 4.266       & 0.018       \\
    ~ & 20 & 3.360       & 0.014       & 3.201       & 0.014       & 3.003       & 0.013       & 2.844       & 0.012       \\
    ~ & 12 & 2.016       & 0.009       & 1.920       & 0.008       & 1.802       & 0.008       & 1.706       & 0.007       \\
    ~ & 6  & 1.008       & 0.004       & 0.960       & 0.004       & 0.901       & 0.004       & 0.853       & 0.004       \\
    ~ & 2  & 0.336       & 0.001       & 0.320       & 0.001       & 0.300       & 0.001       & 0.284       & 0.001       \\
    4 & 42 & 7.056       & 0.030       & 6.722       & 0.028       & 6.306       & 0.027       & 5.972       & 0.025       \\
    ~ & 30 & 5.040       & 0.021       & 4.801       & 0.020       & 4.505       & 0.019       & 4.266       & 0.018       \\
    ~ & 20 & 3.360       & 0.014       & 3.201       & 0.014       & 3.003       & 0.013       & 2.844       & 0.012       \\
    ~ & 12 & 2.016       & 0.009       & 1.920       & 0.008       & 1.802       & 0.008       & 1.706       & 0.007       \\
    ~ & 6  & 1.008       & 0.004       & 0.960       & 0.004       & 0.901       & 0.004       & 0.853       & 0.004       \\
    5 & 56 & 9.407       & 0.040       & 8.962       & 0.038       & 8.409       & 0.036       & 7.963       & 0.034       \\
    ~ & 42 & 7.056       & 0.030       & 6.722       & 0.028       & 6.306       & 0.027       & 5.972       & 0.025       \\
    ~ & 30 & 5.040       & 0.021       & 4.801       & 0.020       & 4.505       & 0.019       & 4.266       & 0.018       \\
    ~ & 20 & 3.360       & 0.014       & 3.201       & 0.014       & 3.003       & 0.013       & 2.844       & 0.012       \\
    ~ & 12 & 2.016       & 0.009       & 1.920       & 0.008       & 1.802       & 0.008       & 1.706       & 0.007       \\
\hline\hline
\end{tabular}
\end{table}

%
%
\clearpage
\section{Character tables}
\FloatBarrier

\begin{table}[ht]
  \caption{%
    Character table of the $C_{2\text{v}}$(M) molecular symmetry group.
    \label{tab:c2vm}
  }
\scalebox{0.9}{%
  \begin{tabular}{@{}l r@{}l r@{}l r@{}l r@{}l r@{}l | r@{}l r@{}l r@{}l r@{}l r@{}l @{}}
  \cline{1-9} \\[-0.4cm]
  \cline{1-9} \\[-0.4cm]
    $C_\mr{2v}$(M)  
    & 
    \multicolumn{2}{c}{$E$} & 
    \multicolumn{2}{c}{(ab)} & 
    \multicolumn{2}{c}{$E^\ast$} &
    \multicolumn{2}{c}{$[(\mr{ab})]^\ast$} \\    
    $n_\mr{cl}$ & 
    \multicolumn{2}{c}{1} & 
    \multicolumn{2}{c}{1} & 
    \multicolumn{2}{c}{1} & 
    \multicolumn{2}{c}{1} \\        
  \cline{1-9} \\[-0.4cm]
  A$_1$ &  &1  &   &1  &   &1  &    &1 \\
  A$_2$ &  &1  &   &1  & $-$&1 & $-$&1 \\  
  B$_1$ &  &1  & $-$&1 & $-$&1 &    &1 \\
  B$_2$ &  &1  & $-$&1 &   &1  & $-$&1 \\
  \cline{1-9} \\[-0.4cm]
  \cline{1-9} \\[-0.4cm]           
  \end{tabular}
}  
\end{table}

\begin{table}[ht]
  \caption{%
    Character table of the $T_\text{d}$(M) molecular symmetry group.
    \label{tab:tdm}
  }
\scalebox{0.9}{%
  \begin{tabular}{@{}l r@{}l r@{}l r@{}l r@{}l r@{}l@{}}
  \cline{1-11} \\[-0.4cm]
  \cline{1-11} \\[-0.4cm]
     $T_\mr{d}$(M) 
    & 
    \multicolumn{2}{c}{$E$} & 
    \multicolumn{2}{c}{(123)} & 
    \multicolumn{2}{c}{(14)(23)} & 
    \multicolumn{2}{c}{[(1423)]$^\ast$} & 
    \multicolumn{2}{c}{[(23)]$^\ast$} \\
    $n_\mr{cl}$ & 
    \multicolumn{2}{c}{1} & 
    \multicolumn{2}{c}{8} & 
    \multicolumn{2}{c}{3} & 
    \multicolumn{2}{c}{6} & 
    \multicolumn{2}{c}{6} \\        
  \cline{1-11} \\[-0.4cm]
  A$_1$ &  &1  &   &1  &   &1  &    &1 &   &1  \\
  A$_2$ &  &1  &   &1  &   &1  & $-$&1 & $-$&1 \\  
  E     &  &2  & $-$&1 &   &2  &   &0  &   &0  \\
  F$_1$ &  &3  &   &0  & $-$&1 &   &1  & $-$&1 \\ 
  F$_2$ &  &3  &   &0  & $-$&1 & $-$&1 &   &1  \\ 
  \cline{1-11} \\[-0.4cm]
  \cline{1-11} \\[-0.4cm]           
  \end{tabular}
}  
\end{table}

\begin{table}[ht]
  \caption{%
    Character table of the molecular symmetry group of CH$_4\cdot$H$_2$O, $G_{48}$ 
    (the table is taken from Ref.~\cite{DoSa94} but with explicit notation for space inversion).
    \label{tab:g48}
  }
  \includegraphics[scale=1.]{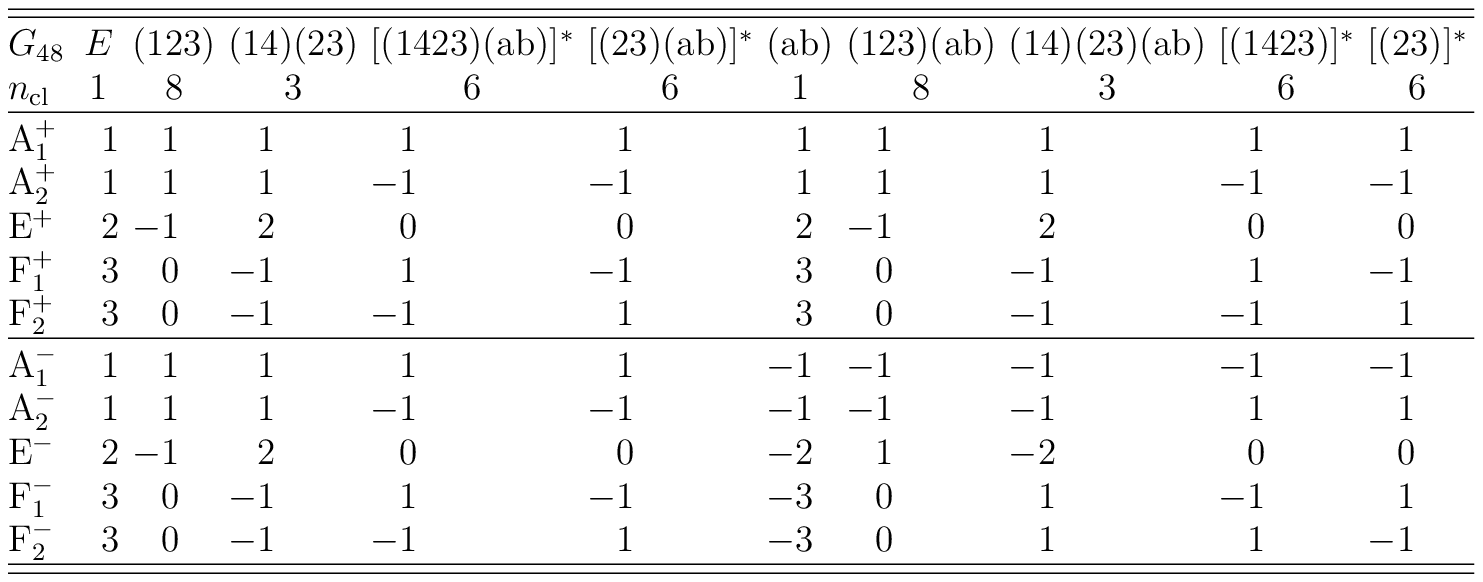}
\end{table}

%
%
\begin{table}
\caption{%
  Characters and irrep decomposition of the 
  $[[\jmet,\jwat_{\ka\kc}]_j,\Lambda]_J$
  coupled-rotor functions which dominate 
  the ZPV(GM) splitting manifold with $J=1$ 
  and $\Gamma^+$ symmetry
  in the $G_{48}$ group
  (see also Figure~3).
  \label{tab:crsymmj1gp}
}
\scalebox{0.8}{%
  \begin{tabular}{@{}l r@{\ \ }l r@{}l r@{}l r@{}l r@{}l  r@{}l r@{}l r@{}l r@{}l r@{}l c @{}}
  \cline{1-22} \\[-0.4cm]
  \cline{1-22} \\[-0.4cm]
    $\Gamma$ & 
    \multicolumn{2}{c}{\small $E$} & 
    \multicolumn{2}{c}{\small (123)} & 
    \multicolumn{2}{c}{\small (14)(23)} & 
    \multicolumn{2}{c}{\small [(1423)(ab)]$^\ast$} & 
    \multicolumn{2}{c}{\small [(23)(ab)]$^\ast$} & 
    \multicolumn{2}{c}{\small (ab)} & 
    \multicolumn{2}{c}{\small (123)(ab)} & 
    \multicolumn{2}{c}{\small (14)(23)(ab)} & 
    \multicolumn{2}{c}{\small [(1423)]$^\ast$} & 
    \multicolumn{2}{c}{\small [(23)]$^\ast$} & 
    Irreps \\    
    \multicolumn{1}{r}{} & 
    \multicolumn{2}{c}{1} & 
    \multicolumn{2}{c}{8} & 
    \multicolumn{2}{c}{3} & 
    \multicolumn{2}{c}{6} & 
    \multicolumn{2}{c}{6} & 
    \multicolumn{2}{c}{1} & 
    \multicolumn{2}{c}{8} & 
    \multicolumn{2}{c}{3} & 
    \multicolumn{2}{c}{6} & 
    \multicolumn{2}{c}{6} & \\        
  \cline{1-22} \\[-0.4cm]
  %
$[[0,0_{00}]_0,1]_1$ & &1 & &1 & &1 & $-$&1 & $-$&1 & &1 & &1 & &1 & $-$&1 & $-$&1 & $\Atwop$ \\
$[[0,1_{11}]_1,0]_1$ & &1 & &1 & &1 & $-$&1 & $-$&1 & &1 & &1 & &1 & $-$&1 & $-$&1 & $\Atwop$ \\
$[[0,1_{11}]_1,1]_1$ & &1 & &1 & &1 & &1 & &1 & &1 & &1 & &1 & &1 & &1 & $\Aonep$ \\
$[[0,1_{11}]_1,2]_1$ & &1 & &1 & &1 & $-$&1 & $-$&1 & &1 & &1 & &1 & $-$&1 & $-$&1 & $\Atwop$ \\
$[[1,0_{00}]_1,0]_1$ & &3 & &0 & $-$&1 & &1 & $-$&1 & &3 & &0 & $-$&1 & &1 & $-$&1 & $\Fonep$ \\
$[[1,0_{00}]_1,1]_1$ & &3 & &0 & $-$&1 & $-$&1 & &1 & &3 & &0 & $-$&1 & $-$&1 & &1 & $\Ftwop$ \\
$[[1,0_{00}]_1,2]_1$ & &3 & &0 & $-$&1 & &1 & $-$&1 & &3 & &0 & $-$&1 & &1 & $-$&1 & $\Fonep$ \\
$[[1,1_{11}]_0,1]_1$ & &3 & &0 & $-$&1 & &1 & $-$&1 & &3 & &0 & $-$&1 & &1 & $-$&1 & $\Fonep$ \\
$[[1,1_{11}]_1,0]_1$ & &3 & &0 & $-$&1 & $-$&1 & &1 & &3 & &0 & $-$&1 & $-$&1 & &1 & $\Ftwop$ \\
$[[1,1_{11}]_1,1]_1$ & &3 & &0 & $-$&1 & &1 & $-$&1 & &3 & &0 & $-$&1 & &1 & $-$&1 & $\Fonep$ \\
$[[1,1_{11}]_1,2]_1$ & &3 & &0 & $-$&1 & $-$&1 & &1 & &3 & &0 & $-$&1 & $-$&1 & &1 & $\Ftwop$ \\
$[[1,1_{11}]_2,1]_1$ & &3 & &0 & $-$&1 & &1 & $-$&1 & &3 & &0 & $-$&1 & &1 & $-$&1 & $\Fonep$ \\
$[[1,1_{11}]_2,2]_1$ & &3 & &0 & $-$&1 & $-$&1 & &1 & &3 & &0 & $-$&1 & $-$&1 & &1 & $\Ftwop$ \\
$[[1,1_{11}]_2,3]_1$ & &3 & &0 & $-$&1 & &1 & $-$&1 & &3 & &0 & $-$&1 & &1 & $-$&1 & $\Fonep$ \\
$[[2,0_{00}]_2,1]_1$ & &5 & $-$&1 & &1 & &1 & $-$&1 & &5 & $-$&1 & &1 & &1 & $-$&1 & $\Ep\oplus\Fonep$ \\
$[[2,0_{00}]_2,2]_1$ & &5 & $-$&1 & &1 & $-$&1 & &1 & &5 & $-$&1 & &1 & $-$&1 & &1 & $\Ep\oplus\Ftwop$ \\
$[[2,0_{00}]_2,3]_1$ & &5 & $-$&1 & &1 & &1 & $-$&1 & &5 & $-$&1 & &1 & &1 & $-$&1 & $\Ep\oplus\Fonep$ \\
$[[2,1_{11}]_1,0]_1$ & &5 & $-$&1 & &1 & &1 & $-$&1 & &5 & $-$&1 & &1 & &1 & $-$&1 & $\Ep\oplus\Fonep$ \\
$[[2,1_{11}]_1,1]_1$ & &5 & $-$&1 & &1 & $-$&1 & &1 & &5 & $-$&1 & &1 & $-$&1 & &1 & $\Ep\oplus\Ftwop$ \\
$[[2,1_{11}]_1,2]_1$ & &5 & $-$&1 & &1 & &1 & $-$&1 & &5 & $-$&1 & &1 & &1 & $-$&1 & $\Ep\oplus\Fonep$ \\
$[[2,1_{11}]_2,1]_1$ & &5 & $-$&1 & &1 & $-$&1 & &1 & &5 & $-$&1 & &1 & $-$&1 & &1 & $\Ep\oplus\Ftwop$ \\
$[[2,1_{11}]_2,2]_1$ & &5 & $-$&1 & &1 & &1 & $-$&1 & &5 & $-$&1 & &1 & &1 & $-$&1 & $\Ep\oplus\Fonep$ \\
$[[2,1_{11}]_2,3]_1$ & &5 & $-$&1 & &1 & $-$&1 & &1 & &5 & $-$&1 & &1 & $-$&1 & &1 & $\Ep\oplus\Ftwop$ \\  
  \cline{1-22} \\[-0.4cm]
  \cline{1-22} \\[-0.4cm]           
  \end{tabular}
}
\end{table}

%
%
\begin{table}
\caption{%
  Characters and irrep decomposition of the 
  $[[\jmet,\jwat_{\ka\kc}]_j,\Lambda]_J$
  coupled-rotor functions which dominate 
  the ZPV(GM) splitting manifold with $J=1$ 
  and $\Gamma^-$ symmetry 
  in the $G_{48}$ group  
  (see also Figure~4).
  \label{tab:crsymmj1gm}
}
\scalebox{0.8}{%
  \begin{tabular}{@{}l r@{\ \ }l r@{}l r@{}l r@{}l r@{}l  r@{}l r@{}l r@{}l r@{}l r@{}l c @{}}
  \cline{1-22} \\[-0.4cm]
  \cline{1-22} \\[-0.4cm]
    $\Gamma$ & 
    \multicolumn{2}{c}{\small $E$} & 
    \multicolumn{2}{c}{\small (123)} & 
    \multicolumn{2}{c}{\small (14)(23)} & 
    \multicolumn{2}{c}{\small [(1423)(ab)]$^\ast$} & 
    \multicolumn{2}{c}{\small [(23)(ab)]$^\ast$} & 
    \multicolumn{2}{c}{\small (ab)} & 
    \multicolumn{2}{c}{\small (123)(ab)} & 
    \multicolumn{2}{c}{\small (14)(23)(ab)} & 
    \multicolumn{2}{c}{\small [(1423)]$^\ast$} & 
    \multicolumn{2}{c}{\small [(23)]$^\ast$} & 
    Irreps \\    
    \multicolumn{1}{r}{} & 
    \multicolumn{2}{c}{1} & 
    \multicolumn{2}{c}{8} & 
    \multicolumn{2}{c}{3} & 
    \multicolumn{2}{c}{6} & 
    \multicolumn{2}{c}{6} & 
    \multicolumn{2}{c}{1} & 
    \multicolumn{2}{c}{8} & 
    \multicolumn{2}{c}{3} & 
    \multicolumn{2}{c}{6} & 
    \multicolumn{2}{c}{6} & \\        
  \cline{1-22} \\[-0.4cm]
  %
$[[0,1_{01}]_1,0]_1$ & &1 & &1 & &1 & &1 & &1 & $-$&1 & $-$&1 & $-$&1 & $-$&1 & $-$&1 & $\Aonem$ \\
$[[0,1_{01}]_1,1]_1$ & &1 & &1 & &1 & $-$&1 & $-$&1 & $-$&1 & $-$&1 & $-$&1 & &1 & &1 & $\Atwom$ \\
$[[0,1_{01}]_1,2]_1$ & &1 & &1 & &1 & &1 & &1 & $-$&1 & $-$&1 & $-$&1 & $-$&1 & $-$&1 & $\Aonem$ \\
$[[0,1_{10}]_1,0]_1$ & &1 & &1 & &1 & $-$&1 & $-$&1 & $-$&1 & $-$&1 & $-$&1 & &1 & &1 & $\Atwom$ \\
$[[0,1_{10}]_1,1]_1$ & &1 & &1 & &1 & &1 & &1 & $-$&1 & $-$&1 & $-$&1 & $-$&1 & $-$&1 & $\Aonem$ \\
$[[0,1_{10}]_1,2]_1$ & &1 & &1 & &1 & $-$&1 & $-$&1 & $-$&1 & $-$&1 & $-$&1 & &1 & &1 & $\Atwom$ \\
$[[1,1_{01}]_0,1]_1$ & &3 & &0 & $-$&1 & $-$&1 & &1 & $-$&3 & &0 & &1 & &1 & $-$&1 & $\Ftwom$ \\
$[[1,1_{01}]_1,0]_1$ & &3 & &0 & $-$&1 & &1 & $-$&1 & $-$&3 & &0 & &1 & $-$&1 & &1 & $\Fonem$\\
$[[1,1_{01}]_1,1]_1$ & &3 & &0 & $-$&1 & $-$&1 & &1 & $-$&3 & &0 & &1 & &1 & $-$&1 & $\Ftwom$ \\
$[[1,1_{01}]_1,2]_1$ & &3 & &0 & $-$&1 & &1 & $-$&1 & $-$&3 & &0 & &1 & $-$&1 & &1 & $\Fonem$\\
$[[1,1_{01}]_2,1]_1$ & &3 & &0 & $-$&1 & $-$&1 & &1 & $-$&3 & &0 & &1 & &1 & $-$&1 & $\Ftwom$ \\
$[[1,1_{01}]_2,2]_1$ & &3 & &0 & $-$&1 & &1 & $-$&1 & $-$&3 & &0 & &1 & $-$&1 & &1 & $\Fonem$\\
$[[1,1_{01}]_2,3]_1$ & &3 & &0 & $-$&1 & $-$&1 & &1 & $-$&3 & &0 & &1 & &1 & $-$&1 & $\Ftwom$ \\
$[[1,1_{10}]_0,1]_1$ & &3 & &0 & $-$&1 & &1 & $-$&1 & $-$&3 & &0 & &1 & $-$&1 & &1 & $\Fonem$\\
$[[1,1_{10}]_1,0]_1$ & &3 & &0 & $-$&1 & $-$&1 & &1 & $-$&3 & &0 & &1 & &1 & $-$&1 & $\Ftwom$ \\
$[[1,1_{10}]_1,1]_1$ & &3 & &0 & $-$&1 & &1 & $-$&1 & $-$&3 & &0 & &1 & $-$&1 & &1 & $\Fonem$\\
$[[1,1_{10}]_1,2]_1$ & &3 & &0 & $-$&1 & $-$&1 & &1 & $-$&3 & &0 & &1 & &1 & $-$&1 & $\Ftwom$ \\
$[[1,1_{10}]_2,1]_1$ & &3 & &0 & $-$&1 & &1 & $-$&1 & $-$&3 & &0 & &1 & $-$&1 & &1 & $\Fonem$\\
$[[1,1_{10}]_2,2]_1$ & &3 & &0 & $-$&1 & $-$&1 & &1 & $-$&3 & &0 & &1 & &1 & $-$&1 & $\Ftwom$ \\
$[[1,1_{10}]_2,3]_1$ & &3 & &0 & $-$&1 & &1 & $-$&1 & $-$&3 & &0 & &1 & $-$&1 & &1 & $\Fonem$\\
$[[2,1_{01}]_1,0]_1$ & &5 & $-$&1 & &1 & $-$&1 & &1 & $-$&5 & &1 & $-$&1 & &1 & $-$&1 & $\Em\oplus\Ftwom$ \\
$[[2,1_{01}]_1,1]_1$ & &5 & $-$&1 & &1 & &1 & $-$&1 & $-$&5 & &1 & $-$&1 & $-$&1 & &1 & $\Em\oplus\Fonem$ \\
$[[2,1_{01}]_1,2]_1$ & &5 & $-$&1 & &1 & $-$&1 & &1 & $-$&5 & &1 & $-$&1 & &1 & $-$&1 & $\Em\oplus\Ftwom$ \\
$[[2,1_{01}]_2,1]_1$ & &5 & $-$&1 & &1 & &1 & $-$&1 & $-$&5 & &1 & $-$&1 & $-$&1 & &1 & $\Em\oplus\Fonem$ \\
$[[2,1_{01}]_2,2]_1$ & &5 & $-$&1 & &1 & $-$&1 & &1 & $-$&5 & &1 & $-$&1 & &1 & $-$&1 & $\Em\oplus\Ftwom$ \\
$[[2,1_{01}]_2,3]_1$ & &5 & $-$&1 & &1 & &1 & $-$&1 & $-$&5 & &1 & $-$&1 & $-$&1 & &1 & $\Em\oplus\Fonem$ \\
$[[2,1_{01}]_3,2]_1$ & &5 & $-$&1 & &1 & $-$&1 & &1 & $-$&5 & &1 & $-$&1 & &1 & $-$&1 & $\Em\oplus\Ftwom$ \\
$[[2,1_{01}]_3,3]_1$ & &5 & $-$&1 & &1 & &1 & $-$&1 & $-$&5 & &1 & $-$&1 & $-$&1 & &1 & $\Em\oplus\Fonem$ \\
$[[2,1_{01}]_3,4]_1$ & &5 & $-$&1 & &1 & $-$&1 & &1 & $-$&5 & &1 & $-$&1 & &1 & $-$&1 & $\Em\oplus\Ftwom$ \\
$[[2,1_{10}]_1,0]_1$ & &5 & $-$&1 & &1 & &1 & $-$&1 & $-$&5 & &1 & $-$&1 & $-$&1 & &1 & $\Em\oplus\Fonem$ \\
$[[2,1_{10}]_1,1]_1$ & &5 & $-$&1 & &1 & $-$&1 & &1 & $-$&5 & &1 & $-$&1 & &1 & $-$&1 & $\Em\oplus\Ftwom$ \\
$[[2,1_{10}]_1,2]_1$ & &5 & $-$&1 & &1 & &1 & $-$&1 & $-$&5 & &1 & $-$&1 & $-$&1 & &1 & $\Em\oplus\Fonem$ \\
$[[2,1_{10}]_2,1]_1$ & &5 & $-$&1 & &1 & $-$&1 & &1 & $-$&5 & &1 & $-$&1 & &1 & $-$&1 & $\Em\oplus\Ftwom$ \\
$[[2,1_{10}]_2,2]_1$ & &5 & $-$&1 & &1 & &1 & $-$&1 & $-$&5 & &1 & $-$&1 & $-$&1 & &1 & $\Em\oplus\Fonem$ \\
$[[2,1_{10}]_2,3]_1$ & &5 & $-$&1 & &1 & $-$&1 & &1 & $-$&5 & &1 & $-$&1 & &1 & $-$&1 & $\Em\oplus\Ftwom$ \\
$[[2,1_{10}]_3,2]_1$ & &5 & $-$&1 & &1 & &1 & $-$&1 & $-$&5 & &1 & $-$&1 & $-$&1 & &1 & $\Em\oplus\Fonem$ \\
$[[2,1_{10}]_3,3]_1$ & &5 & $-$&1 & &1 & $-$&1 & &1 & $-$&5 & &1 & $-$&1 & &1 & $-$&1 & $\Em\oplus\Ftwom$ \\
$[[2,1_{10}]_3,4]_1$ & &5 & $-$&1 & &1 & &1 & $-$&1 & $-$&5 & &1 & $-$&1 & $-$&1 & &1 & $\Em\oplus\Fonem$ \\  
  \cline{1-22} \\[-0.4cm]
  \cline{1-22} \\[-0.4cm]         
  \end{tabular}
}
\end{table}

%
%
\begin{table}
\caption{%
  Characters and irrep decomposition of the 
  $[[\jmet,\jwat_{\ka\kc}]_j,\Lambda]_J$
  coupled-rotor functions which dominate 
  the ZPV(GM) splitting manifold with $J=2$ 
  and $\Gamma^+$ symmetry 
  in the $G_{48}$ group  
  (see also Figure~3).  
  \label{tab:crsymmj2gp}
}
\scalebox{0.8}{%
  \begin{tabular}{@{}l r@{\ \ }l r@{}l r@{}l r@{}l r@{}l  r@{}l r@{}l r@{}l r@{}l r@{}l c @{}}
  \cline{1-22} \\[-0.4cm]
  \cline{1-22} \\[-0.4cm]
    $\Gamma$ & 
    \multicolumn{2}{c}{\small $E$} & 
    \multicolumn{2}{c}{\small (123)} & 
    \multicolumn{2}{c}{\small (14)(23)} & 
    \multicolumn{2}{c}{\small [(1423)(ab)]$^\ast$} & 
    \multicolumn{2}{c}{\small [(23)(ab)]$^\ast$} & 
    \multicolumn{2}{c}{\small (ab)} & 
    \multicolumn{2}{c}{\small (123)(ab)} & 
    \multicolumn{2}{c}{\small (14)(23)(ab)} & 
    \multicolumn{2}{c}{\small [(1423)]$^\ast$} & 
    \multicolumn{2}{c}{\small [(23)]$^\ast$} & 
    Irreps \\    
    \multicolumn{1}{r}{} & 
    \multicolumn{2}{c}{1} & 
    \multicolumn{2}{c}{8} & 
    \multicolumn{2}{c}{3} & 
    \multicolumn{2}{c}{6} & 
    \multicolumn{2}{c}{6} & 
    \multicolumn{2}{c}{1} & 
    \multicolumn{2}{c}{8} & 
    \multicolumn{2}{c}{3} & 
    \multicolumn{2}{c}{6} & 
    \multicolumn{2}{c}{6} & \\        
  \cline{1-22} \\[-0.4cm]
  %
$[[0,0_{00}]_0,2]_2$ & &1 & &1 & &1 & &1 & &1 & &1 & &1 & &1 & &1 & &1 & $\Aonep$ \\
$[[0,1_{11}]_1,1]_2$ & &1 & &1 & &1 & &1 & &1 & &1 & &1 & &1 & &1 & &1 & $\Aonep$ \\
$[[0,1_{11}]_1,2]_2$ & &1 & &1 & &1 & $-$&1 & $-$&1 & &1 & &1 & &1 & $-$&1 & $-$&1 & $\Atwop$ \\
$[[0,1_{11}]_1,3]_2$ & &1 & &1 & &1 & &1 & &1 & &1 & &1 & &1 & &1 & &1 & $\Aonep$ \\
$[[1,0_{00}]_1,1]_2$ & &3 & &0 & $-$&1 & $-$&1 & &1 & &3 & &0 & $-$&1 & $-$&1 & &1 & $\Ftwop$ \\
$[[1,0_{00}]_1,2]_2$ & &3 & &0 & $-$&1 & &1 & $-$&1 & &3 & &0 & $-$&1 & &1 & $-$&1 & $\Fonep$ \\
$[[1,0_{00}]_1,3]_2$ & &3 & &0 & $-$&1 & $-$&1 & &1 & &3 & &0 & $-$&1 & $-$&1 & &1 & $\Ftwop$ \\
$[[1,1_{11}]_0,2]_2$ & &3 & &0 & $-$&1 & $-$&1 & &1 & &3 & &0 & $-$&1 & $-$&1 & &1 & $\Ftwop$ \\
$[[1,1_{11}]_1,1]_2$ & &3 & &0 & $-$&1 & &1 & $-$&1 & &3 & &0 & $-$&1 & &1 & $-$&1 & $\Fonep$ \\
$[[1,1_{11}]_1,2]_2$ & &3 & &0 & $-$&1 & $-$&1 & &1 & &3 & &0 & $-$&1 & $-$&1 & &1 & $\Ftwop$ \\
$[[1,1_{11}]_1,3]_2$ & &3 & &0 & $-$&1 & &1 & $-$&1 & &3 & &0 & $-$&1 & &1 & $-$&1 & $\Fonep$ \\
$[[2,0_{00}]_2,0]_2$ & &5 & $-$&1 & &1 & $-$&1 & &1 & &5 & $-$&1 & &1 & $-$&1 & &1 & $\Ep\oplus\Ftwop$ \\
$[[2,0_{00}]_2,1]_2$ & &5 & $-$&1 & &1 & &1 & $-$&1 & &5 & $-$&1 & &1 & &1 & $-$&1 & $\Ep\oplus\Fonep$ \\
$[[2,0_{00}]_2,2]_2$ & &5 & $-$&1 & &1 & $-$&1 & &1 & &5 & $-$&1 & &1 & $-$&1 & &1 & $\Ep\oplus\Ftwop$ \\
$[[2,0_{00}]_2,3]_2$ & &5 & $-$&1 & &1 & &1 & $-$&1 & &5 & $-$&1 & &1 & &1 & $-$&1 & $\Ep\oplus\Fonep$ \\
$[[2,0_{00}]_2,4]_2$ & &5 & $-$&1 & &1 & $-$&1 & &1 & &5 & $-$&1 & &1 & $-$&1 & &1 & $\Ep\oplus\Ftwop$ \\
$[[2,1_{11}]_1,1]_2$ & &5 & $-$&1 & &1 & $-$&1 & &1 & &5 & $-$&1 & &1 & $-$&1 & &1 & $\Ep\oplus\Ftwop$ \\
$[[2,1_{11}]_1,2]_2$ & &5 & $-$&1 & &1 & &1 & $-$&1 & &5 & $-$&1 & &1 & &1 & $-$&1 & $\Ep\oplus\Fonep$ \\
$[[2,1_{11}]_1,3]_2$ & &5 & $-$&1 & &1 & $-$&1 & &1 & &5 & $-$&1 & &1 & $-$&1 & &1 & $\Ep\oplus\Ftwop$ \\
$[[2,1_{11}]_2,0]_2$ & &5 & $-$&1 & &1 & &1 & $-$&1 & &5 & $-$&1 & &1 & &1 & $-$&1 & $\Ep\oplus\Fonep$ \\
$[[2,1_{11}]_2,1]_2$ & &5 & $-$&1 & &1 & $-$&1 & &1 & &5 & $-$&1 & &1 & $-$&1 & &1 & $\Ep\oplus\Ftwop$ \\
$[[2,1_{11}]_2,2]_2$ & &5 & $-$&1 & &1 & &1 & $-$&1 & &5 & $-$&1 & &1 & &1 & $-$&1 & $\Ep\oplus\Fonep$ \\
$[[2,1_{11}]_2,3]_2$ & &5 & $-$&1 & &1 & $-$&1 & &1 & &5 & $-$&1 & &1 & $-$&1 & &1 & $\Ep\oplus\Ftwop$ \\
$[[2,1_{11}]_2,4]_2$ & &5 & $-$&1 & &1 & &1 & $-$&1 & &5 & $-$&1 & &1 & &1 & $-$&1 & $\Ep\oplus\Fonep$ \\
$[[2,1_{11}]_3,1]_2$ & &5 & $-$&1 & &1 & $-$&1 & &1 & &5 & $-$&1 & &1 & $-$&1 & &1 & $\Ep\oplus\Ftwop$ \\
$[[2,1_{11}]_3,2]_2$ & &5 & $-$&1 & &1 & &1 & $-$&1 & &5 & $-$&1 & &1 & &1 & $-$&1 & $\Ep\oplus\Fonep$ \\
$[[2,1_{11}]_3,3]_2$ & &5 & $-$&1 & &1 & $-$&1 & &1 & &5 & $-$&1 & &1 & $-$&1 & &1 & $\Ep\oplus\Ftwop$ \\
$[[2,1_{11}]_3,4]_2$ & &5 & $-$&1 & &1 & &1 & $-$&1 & &5 & $-$&1 & &1 & &1 & $-$&1 & $\Ep\oplus\Fonep$ \\
$[[2,1_{11}]_3,5]_2$ & &5 & $-$&1 & &1 & $-$&1 & &1 & &5 & $-$&1 & &1 & $-$&1 & &1 & $\Ep\oplus\Ftwop$ \\  
  \cline{1-22} \\[-0.4cm]
  \cline{1-22} \\[-0.4cm]           
  \end{tabular}
}
\end{table}

%
%
\begin{table}
\caption{%
  Characters and irrep decomposition of the 
  $[[\jmet,\jwat_{\ka\kc}]_j,\Lambda]_J$
  coupled-rotor functions which dominate 
  the ZPV(GM) splitting manifold with $J=2$ 
  and $\Gamma^-$ symmetry 
  in the $G_{48}$ group  
  (see also Figure~4).
  \label{tab:crsymmj2gm}
}
\scalebox{0.8}{%
  \begin{tabular}{@{}l r@{\ \ }l r@{}l r@{}l r@{}l r@{}l  r@{}l r@{}l r@{}l r@{}l r@{}l c @{}}
  \cline{1-22} \\[-0.4cm]
  \cline{1-22} \\[-0.4cm]
    $\Gamma$ & 
    \multicolumn{2}{c}{\small $E$} & 
    \multicolumn{2}{c}{\small (123)} & 
    \multicolumn{2}{c}{\small (14)(23)} & 
    \multicolumn{2}{c}{\small [(1423)(ab)]$^\ast$} & 
    \multicolumn{2}{c}{\small [(23)(ab)]$^\ast$} & 
    \multicolumn{2}{c}{\small (ab)} & 
    \multicolumn{2}{c}{\small (123)(ab)} & 
    \multicolumn{2}{c}{\small (14)(23)(ab)} & 
    \multicolumn{2}{c}{\small [(1423)]$^\ast$} & 
    \multicolumn{2}{c}{\small [(23)]$^\ast$} & 
    Irreps \\    
    \multicolumn{1}{r}{} & 
    \multicolumn{2}{c}{1} & 
    \multicolumn{2}{c}{8} & 
    \multicolumn{2}{c}{3} & 
    \multicolumn{2}{c}{6} & 
    \multicolumn{2}{c}{6} & 
    \multicolumn{2}{c}{1} & 
    \multicolumn{2}{c}{8} & 
    \multicolumn{2}{c}{3} & 
    \multicolumn{2}{c}{6} & 
    \multicolumn{2}{c}{6} & \\        
  \cline{1-22} \\[-0.4cm]
  %
$[[0,1_{01}]_1,1]_2$ & &1 & &1 & &1 & $-$&1 & $-$&1 & $-$&1 & $-$&1 & $-$&1 & &1 & &1 & $\Atwom$ \\
$[[0,1_{01}]_1,2]_2$ & &1 & &1 & &1 & &1 & &1 & $-$&1 & $-$&1 & $-$&1 & $-$&1 & $-$&1 & $\Aonem$ \\
$[[0,1_{01}]_1,3]_2$ & &1 & &1 & &1 & $-$&1 & $-$&1 & $-$&1 & $-$&1 & $-$&1 & &1 & &1 & $\Atwom$ \\
$[[0,1_{10}]_1,1]_2$ & &1 & &1 & &1 & &1 & &1 & $-$&1 & $-$&1 & $-$&1 & $-$&1 & $-$&1 & $\Aonem$ \\
$[[0,1_{10}]_1,2]_2$ & &1 & &1 & &1 & $-$&1 & $-$&1 & $-$&1 & $-$&1 & $-$&1 & &1 & &1 & $\Atwom$ \\
$[[0,1_{10}]_1,3]_2$ & &1 & &1 & &1 & &1 & &1 & $-$&1 & $-$&1 & $-$&1 & $-$&1 & $-$&1 & $\Aonem$ \\
$[[1,1_{01}]_0,2]_2$ & &3 & &0 & $-$&1 & &1 & $-$&1 & $-$&3 & &0 & &1 & $-$&1 & &1 & $\Fonem$ \\
$[[1,1_{01}]_1,1]_2$ & &3 & &0 & $-$&1 & $-$&1 & &1 & $-$&3 & &0 & &1 & &1 & $-$&1 & $\Ftwom$ \\
$[[1,1_{01}]_1,2]_2$ & &3 & &0 & $-$&1 & &1 & $-$&1 & $-$&3 & &0 & &1 & $-$&1 & &1 & $\Fonem$ \\
$[[1,1_{01}]_1,3]_2$ & &3 & &0 & $-$&1 & $-$&1 & &1 & $-$&3 & &0 & &1 & &1 & $-$&1 & $\Ftwom$ \\
$[[1,1_{01}]_2,0]_2$ & &3 & &0 & $-$&1 & &1 & $-$&1 & $-$&3 & &0 & &1 & $-$&1 & &1 & $\Fonem$ \\
$[[1,1_{01}]_2,1]_2$ & &3 & &0 & $-$&1 & $-$&1 & &1 & $-$&3 & &0 & &1 & &1 & $-$&1 & $\Ftwom$ \\
$[[1,1_{01}]_2,2]_2$ & &3 & &0 & $-$&1 & &1 & $-$&1 & $-$&3 & &0 & &1 & $-$&1 & &1 & $\Fonem$ \\
$[[1,1_{01}]_2,3]_2$ & &3 & &0 & $-$&1 & $-$&1 & &1 & $-$&3 & &0 & &1 & &1 & $-$&1 & $\Ftwom$ \\
$[[1,1_{01}]_2,4]_2$ & &3 & &0 & $-$&1 & &1 & $-$&1 & $-$&3 & &0 & &1 & $-$&1 & &1 & $\Fonem$ \\
$[[1,1_{10}]_0,2]_2$ & &3 & &0 & $-$&1 & $-$&1 & &1 & $-$&3 & &0 & &1 & &1 & $-$&1 & $\Ftwom$ \\
$[[1,1_{10}]_1,1]_2$ & &3 & &0 & $-$&1 & &1 & $-$&1 & $-$&3 & &0 & &1 & $-$&1 & &1 & $\Fonem$ \\
$[[1,1_{10}]_1,2]_2$ & &3 & &0 & $-$&1 & $-$&1 & &1 & $-$&3 & &0 & &1 & &1 & $-$&1 & $\Ftwom$ \\
$[[1,1_{10}]_1,3]_2$ & &3 & &0 & $-$&1 & &1 & $-$&1 & $-$&3 & &0 & &1 & $-$&1 & &1 & $\Fonem$ \\
$[[1,1_{10}]_2,0]_2$ & &3 & &0 & $-$&1 & $-$&1 & &1 & $-$&3 & &0 & &1 & &1 & $-$&1 & $\Ftwom$ \\
$[[1,1_{10}]_2,1]_2$ & &3 & &0 & $-$&1 & &1 & $-$&1 & $-$&3 & &0 & &1 & $-$&1 & &1 & $\Fonem$ \\
$[[1,1_{10}]_2,2]_2$ & &3 & &0 & $-$&1 & $-$&1 & &1 & $-$&3 & &0 & &1 & &1 & $-$&1 & $\Ftwom$ \\
$[[1,1_{10}]_2,3]_2$ & &3 & &0 & $-$&1 & &1 & $-$&1 & $-$&3 & &0 & &1 & $-$&1 & &1 & $\Fonem$ \\
$[[1,1_{10}]_2,4]_2$ & &3 & &0 & $-$&1 & $-$&1 & &1 & $-$&3 & &0 & &1 & &1 & $-$&1 & $\Ftwom$ \\
$[[2,1_{01}]_1,1]_2$ & &5 & $-$&1 & &1 & &1 & $-$&1 & $-$&5 & &1 & $-$&1 & $-$&1 & &1 & $\Em\oplus\Fonem$ \\
$[[2,1_{01}]_1,2]_2$ & &5 & $-$&1 & &1 & $-$&1 & &1 & $-$&5 & &1 & $-$&1 & &1 & $-$&1 & $\Em\oplus\Ftwom$ \\
$[[2,1_{01}]_1,3]_2$ & &5 & $-$&1 & &1 & &1 & $-$&1 & $-$&5 & &1 & $-$&1 & $-$&1 & &1 & $\Em\oplus\Fonem$ \\
$[[2,1_{01}]_2,0]_2$ & &5 & $-$&1 & &1 & $-$&1 & &1 & $-$&5 & &1 & $-$&1 & &1 & $-$&1 & $\Em\oplus\Ftwom$ \\
$[[2,1_{01}]_2,1]_2$ & &5 & $-$&1 & &1 & &1 & $-$&1 & $-$&5 & &1 & $-$&1 & $-$&1 & &1 & $\Em\oplus\Fonem$ \\
$[[2,1_{01}]_2,2]_2$ & &5 & $-$&1 & &1 & $-$&1 & &1 & $-$&5 & &1 & $-$&1 & &1 & $-$&1 & $\Em\oplus\Ftwom$ \\
$[[2,1_{01}]_2,3]_2$ & &5 & $-$&1 & &1 & &1 & $-$&1 & $-$&5 & &1 & $-$&1 & $-$&1 & &1 & $\Em\oplus\Fonem$ \\
$[[2,1_{01}]_2,4]_2$ & &5 & $-$&1 & &1 & $-$&1 & &1 & $-$&5 & &1 & $-$&1 & &1 & $-$&1 & $\Em\oplus\Ftwom$ \\
$[[2,1_{01}]_3,1]_2$ & &5 & $-$&1 & &1 & &1 & $-$&1 & $-$&5 & &1 & $-$&1 & $-$&1 & &1 & $\Em\oplus\Fonem$ \\
$[[2,1_{01}]_3,2]_2$ & &5 & $-$&1 & &1 & $-$&1 & &1 & $-$&5 & &1 & $-$&1 & &1 & $-$&1 & $\Em\oplus\Ftwom$ \\
$[[2,1_{01}]_3,3]_2$ & &5 & $-$&1 & &1 & &1 & $-$&1 & $-$&5 & &1 & $-$&1 & $-$&1 & &1 & $\Em\oplus\Fonem$ \\
$[[2,1_{01}]_3,4]_2$ & &5 & $-$&1 & &1 & $-$&1 & &1 & $-$&5 & &1 & $-$&1 & &1 & $-$&1 & $\Em\oplus\Ftwom$ \\
$[[2,1_{01}]_3,5]_2$ & &5 & $-$&1 & &1 & &1 & $-$&1 & $-$&5 & &1 & $-$&1 & $-$&1 & &1 & $\Em\oplus\Fonem$ \\
$[[2,1_{10}]_1,1]_2$ & &5 & $-$&1 & &1 & $-$&1 & &1 & $-$&5 & &1 & $-$&1 & &1 & $-$&1 & $\Em\oplus\Ftwom$ \\
$[[2,1_{10}]_1,2]_2$ & &5 & $-$&1 & &1 & &1 & $-$&1 & $-$&5 & &1 & $-$&1 & $-$&1 & &1 & $\Em\oplus\Fonem$ \\
$[[2,1_{10}]_1,3]_2$ & &5 & $-$&1 & &1 & $-$&1 & &1 & $-$&5 & &1 & $-$&1 & &1 & $-$&1 & $\Em\oplus\Ftwom$ \\
$[[2,1_{10}]_2,0]_2$ & &5 & $-$&1 & &1 & &1 & $-$&1 & $-$&5 & &1 & $-$&1 & $-$&1 & &1 & $\Em\oplus\Fonem$ \\
$[[2,1_{10}]_2,1]_2$ & &5 & $-$&1 & &1 & $-$&1 & &1 & $-$&5 & &1 & $-$&1 & &1 & $-$&1 & $\Em\oplus\Ftwom$ \\
$[[2,1_{10}]_2,2]_2$ & &5 & $-$&1 & &1 & &1 & $-$&1 & $-$&5 & &1 & $-$&1 & $-$&1 & &1 & $\Em\oplus\Fonem$ \\
$[[2,1_{10}]_2,3]_2$ & &5 & $-$&1 & &1 & $-$&1 & &1 & $-$&5 & &1 & $-$&1 & &1 & $-$&1 & $\Em\oplus\Ftwom$ \\
$[[2,1_{10}]_2,4]_2$ & &5 & $-$&1 & &1 & &1 & $-$&1 & $-$&5 & &1 & $-$&1 & $-$&1 & &1 & $\Em\oplus\Fonem$ \\
$[[2,1_{10}]_3,1]_2$ & &5 & $-$&1 & &1 & $-$&1 & &1 & $-$&5 & &1 & $-$&1 & &1 & $-$&1 & $\Em\oplus\Ftwom$ \\
$[[2,1_{10}]_3,2]_2$ & &5 & $-$&1 & &1 & &1 & $-$&1 & $-$&5 & &1 & $-$&1 & $-$&1 & &1 & $\Em\oplus\Fonem$ \\
$[[2,1_{10}]_3,3]_2$ & &5 & $-$&1 & &1 & $-$&1 & &1 & $-$&5 & &1 & $-$&1 & &1 & $-$&1 & $\Em\oplus\Ftwom$ \\
$[[2,1_{10}]_3,4]_2$ & &5 & $-$&1 & &1 & &1 & $-$&1 & $-$&5 & &1 & $-$&1 & $-$&1 & &1 & $\Em\oplus\Fonem$ \\
$[[2,1_{10}]_3,5]_2$ & &5 & $-$&1 & &1 & $-$&1 & &1 & $-$&5 & &1 & $-$&1 & &1 & $-$&1 & $\Em\oplus\Ftwom$ \\  
  \cline{1-22} \\[-0.4cm]
  \cline{1-22} \\[-0.4cm]           
  \end{tabular}
}
\end{table}

\FloatBarrier